% ...

\documentclass[preprint]{aastex}

\newcommand{\tnm}{\tablenotemark}
\newcommand{\tnt}{\tablenotetext}
\newcommand{\ch}{\colhead}
\newcommand{\mc}{\multicolumn}

\newcommand{\Av}{$A_V$}
\newcommand{\Msolar}{$M_\odot$}

\newcommand{\lkha}{LkH$\alpha$~}

\newcommand{\Spitzer}{\textit{Spitzer}}
\newcommand{\Herschel}{\textit{Herschel}}
\newcommand{\getsources}{\textit{getsources}}

\newcommand{\co}{$^{12}$CO}

\newcommand{\kms}{km/s}

\defcitealias{Gutermuthetal2009}{G09}
\defcitealias{BroekhovenFieneetal2014a}{BF14}
\defcitealias{Harveyetal2013}{H13}

\usepackage{graphicx}
\usepackage{epsfig}
\usepackage{epstopdf}
\usepackage{natbib}
\usepackage{color}
\usepackage{float}
\usepackage[caption=false]{subfig}
\usepackage{ulem}
\usepackage{xcolor}

\definecolor{edited}{RGB}{128,128,128}
\definecolor{darkgreen}{RGB}{0,128,0}

\shorttitle{AMC with SCUBA-2}
\shortauthors{Broekhoven-Fiene, Matthews et al.}

\begin{document}

\title{The JCMT Gould Belt Survey: A First Look at the Auriga--California Molecular Cloud with SCUBA-2}

\author{
H. Broekhoven-Fiene\altaffilmark{1}, 
B.C. Matthews\altaffilmark{2, 1}, 
P. Harvey\altaffilmark{3}, 
H. Kirk\altaffilmark{2}, 
M. Chen\altaffilmark{1}, 
M.J. Currie\altaffilmark{7}, 
K. Pattle\altaffilmark{4}, 
J. Lane\altaffilmark{2},  
J. Buckle\altaffilmark{5, 6}, 
J. Di Francesco\altaffilmark{2, 1},  
E. Drabek-Maunder\altaffilmark{20},  
D. Johnstone\altaffilmark{7, 2, 1}, 
D.S. Berry\altaffilmark{7}, 
M. Fich\altaffilmark{8}, 
J. Hatchell\altaffilmark{9},  
T. Jenness\altaffilmark{7, 10}, 
J.C. Mottram\altaffilmark{11}, 
D. Nutter\altaffilmark{12},  
J.E. Pineda\altaffilmark{13, 14, 15}, 
C. Quinn\altaffilmark{12}, 
C. Salji\altaffilmark{5, 6}, 
S. Tisi\altaffilmark{8}, 
M.R. Hogerheijde\altaffilmark{11}, 
D. Ward-Thompson\altaffilmark{4}, 
P. Bastien\altaffilmark{16}, 
D. Bresnahan\altaffilmark{4}, 
H. Butner\altaffilmark{17},  
A. Chrysostomou\altaffilmark{18}, 
S. Coude\altaffilmark{16}, 
C.J. Davis\altaffilmark{19}, 
A. Duarte-Cabral\altaffilmark{9}, 
J. Fiege\altaffilmark{21}, 
P. Friberg\altaffilmark{7}, 
R. Friesen\altaffilmark{22}, 
G.A. Fuller\altaffilmark{14},  
S. Graves\altaffilmark{7}, 
J. Greaves\altaffilmark{23}, 
J. Gregson\altaffilmark{24, 25}, 
W. Holland\altaffilmark{26, 27},  
G. Joncas\altaffilmark{28}, 
J.M. Kirk\altaffilmark{4}, 
L.B.G. Knee\altaffilmark{2}, 
S. Mairs\altaffilmark{1}, 
K. Marsh\altaffilmark{12}, 
G. Moriarty-Schieven\altaffilmark{2},  
C. Mowat\altaffilmark{9}, 
J. Rawlings\altaffilmark{29}, 
J. Richer\altaffilmark{5, 6}, 
D. Robertson\altaffilmark{30}, 
E. Rosolowsky\altaffilmark{31},  
D. Rumble\altaffilmark{9}, 
S. Sadavoy\altaffilmark{32}, 
H. Thomas\altaffilmark{7}, 
N. Tothill\altaffilmark{33}, 
S. Viti\altaffilmark{29},  
G.J. White\altaffilmark{24, 25}, 
C.D. Wilson\altaffilmark{30}, 
J. Wouterloot\altaffilmark{7}, 
J. Yates\altaffilmark{29}, 
M. Zhu\altaffilmark{34}}
\altaffiltext{1}{Department of Physics and Astronomy, University of Victoria, Victoria, BC, V8P 1A1, Canada}
\altaffiltext{2}{NRC Herzberg Astronomy and Astrophysics, 5071 West Saanich Rd, Victoria, BC, V9E 2E7, Canada}
\altaffiltext{3}{Astronomy Department, University of Texas at Austin, 1 University Station C1400, Austin, TX 78712-0259, USA}
\altaffiltext{4}{Jeremiah Horrocks Institute, University of Central Lancashire, Preston, Lancashire, PR1 2HE, UK}
\altaffiltext{5}{Astrophysics Group, Cavendish Laboratory, J J Thomson Avenue, Cambridge, CB3 0HE, UK}
\altaffiltext{6}{Kavli Institute for Cosmology, Institute of Astronomy, University of Cambridge, Madingley Road, Cambridge, CB3 0HA, UK}
\altaffiltext{7}{Joint Astronomy Centre, 660 N. A`oh\={o}k\={u} Place, University Park, Hilo, Hawaii 96720, USA}
\altaffiltext{8}{Department of Physics and Astronomy, University of Waterloo, Waterloo, Ontario, N2L 3G1, Canada}
\altaffiltext{9}{Physics and Astronomy, University of Exeter, Stocker Road, Exeter EX4 4QL, UK}
\altaffiltext{10}{LSST Project Office, 933 N. Cherry Ave, Tucson, AZ 85719, USA}
\altaffiltext{11}{Leiden Observatory, Leiden University, PO Box 9513, 2300 RA Leiden, The Netherlands}
\altaffiltext{12}{School of Physics and Astronomy, Cardiff University, The Parade, Cardiff, CF24 3AA, UK}
\altaffiltext{13}{European Southern Observatory (ESO), Garching, Germany}
\altaffiltext{14}{Jodrell Bank Centre for Astrophysics, Alan Turing Building, School of Physics and Astronomy, University of Manchester, Oxford Road, Manchester, M13 9PL, UK}
\altaffiltext{15}{Current address: Institute for Astronomy, ETH Zurich, Wolfgang-Pauli-Strasse 27, CH-8093 Zurich, Switzerland}
\altaffiltext{16}{Universit\'e de Montr\'eal, Centre de Recherche en Astrophysique du Qu\'ebec et d\'epartement de physique, C.P. 6128, succ. centre-ville, Montr\'eal, QC, H3C 3J7, Canada}
\altaffiltext{17}{James Madison University, Harrisonburg, Virginia 22807, USA}
\altaffiltext{18}{School of Physics, Astronomy \& Mathematics, University of Hertfordshire, College Lane, Hatfield, HERTS AL10 9AB, UK}
\altaffiltext{19}{Astrophysics Research Institute, Liverpool John Moores University, Egerton Warf, Birkenhead, CH41 1LD, UK}
\altaffiltext{20}{Imperial College London, Blackett Laboratory, Prince Consort Rd, London SW7 2BB, UK}
\altaffiltext{21}{Dept of Physics \& Astronomy, University of Manitoba, Winnipeg, Manitoba, R3T 2N2 Canada}
\altaffiltext{22}{Dunlap Institute for Astronomy \& Astrophysics, University of Toronto, 50 St. George St., Toronto ON M5S 3H4 Canada}
\altaffiltext{23}{Physics \& Astronomy, University of St Andrews, North Haugh, St Andrews, Fife KY16 9SS, UK}
\altaffiltext{24}{Dept. of Physical Sciences, The Open University, Milton Keynes MK7 6AA, UK}
\altaffiltext{25}{The Rutherford Appleton Laboratory, Chilton, Didcot, OX11 0NL, UK.}
\altaffiltext{26}{UK Astronomy Technology Centre, Royal Observatory, Blackford Hill, Edinburgh EH9 3HJ, UK}
\altaffiltext{27}{Institute for Astronomy, Royal Observatory, University of Edinburgh, Blackford Hill, Edinburgh EH9 3HJ, UK}
\altaffiltext{28}{Centre de recherche en astrophysique du Qu\'ebec et D\'epartement de physique, de g\'enie physique et d'optique, Universit\'e Laval, 1045 avenue de la m\'edecine, Qu\'ebec, G1V 0A6, Canada}
\altaffiltext{29}{Department of Physics and Astronomy, UCL, Gower St, London, WC1E 6BT, UK}
\altaffiltext{30}{Department of Physics and Astronomy, McMaster University, Hamilton, ON, L8S 4M1, Canada}
\altaffiltext{31}{Department of Physics, University of Alberta, Edmonton, AB T6G 2E1, Canada}
\altaffiltext{32}{Max Planck Institute for Astronomy, K\"{o}nigstuhl 17, D-69117 Heidelberg, Germany}
\altaffiltext{33}{University of Western Sydney, Locked Bag 1797, Penrith NSW 2751, Australia}
\altaffiltext{34}{National Astronomical Observatory of China, 20A Datun Road, Chaoyang District, Beijing 100012, China}

\begin{abstract}
We present 850 and 450~\micron\ observations of the dense regions within the Auriga--California molecular cloud using SCUBA-2 as part of the JCMT Gould Belt Legacy Survey to identify candidate protostellar objects, measure the masses of their circumstellar material (disk and envelope), and compare the star formation to that in the Orion A molecular cloud. We identify 59 candidate protostars based on the presence of compact submillimeter emission, complementing these observations with existing \Herschel/SPIRE maps. Of our candidate protostars, 24 are associated with young stellar objects (YSOs) in the \Spitzer\ and \Herschel/PACS catalogs of 166 and 60 YSOs, respectively (177 unique), confirming their protostellar nature. The remaining 35 candidate protostars are in regions, particularly around \lkha 101, where the background cloud emission is too bright to verify or rule out the presence of the compact 70~\micron\ emission that is expected for a protostellar source. We keep these candidate protostars in our sample but note that they may indeed be prestellar in nature.  Our observations are sensitive to the high end of the mass distribution in Auriga--Cal. We find that the disparity between the richness of infrared star forming objects in Orion A and the sparsity in Auriga--Cal extends to the submillimeter, suggesting that the relative star formation rates have not varied over the Class II lifetime and that Auriga--Cal will maintain a lower star formation efficiency.  

\end{abstract}

\section{Introduction}
\label{sec:intro}

The Auriga--California molecular cloud (Auriga--Cal) is a nearby (450 $\pm$ 23 pc: \citealt{Ladaetal2009}) giant molecular cloud notable for its relatively quiescent star formation, in contrast to the Orion A molecular cloud (Orion A). Auriga--Cal was first identified as a contiguous cloud and located in the Gould Belt by \cite{Ladaetal2009}, who also noted that despite Auriga--Cal and Orion A sharing a similar filamentary morphology, as well as similar mass ($\sim10^5$\Msolar), spatial scale (80 pc), and distance (i.e., similar physical characteristics and no drastic observational bias), Auriga--Cal appeared to have much less ongoing star-formation. Lada et al. attributed this deficit of star formation to the lower mass of the cloud at high density. (Orion A North has $\sim$ 8 times more mass at $A_K >$ 1 than Auriga--Cal.) The \Spitzer\ Survey of Interstellar Clouds in the Gould Belt (PI: L. Allen) extended the area of Auriga--Cal surveyed by \Spitzer\ beyond just the young stellar cluster region NGC 1529 around \lkha 101 (observed by \citealt{Gutermuthetal2009}) and confirmed this deficit with a census of the young stellar object (YSO) population throughout the cloud. This census showed that Auriga--Cal contains 15-20 times fewer \Spitzer-identified YSOs than Orion A  (\citealt{BroekhovenFieneetal2014a}), comparable to the ratio of high-density material between the two clouds.  Combined with Auriga--Cal's single early-B star, \lkha 101, in contrast to Orion A's dozens of OB stars, star formation in Auriga--Cal appears more like that in lower-mass clouds like Taurus and Ophiuchus. The classification of the YSOs reveals a high fraction of Class I and F (flat spectrum) YSOs  (associated with early, short-lived stages of star formation), suggesting that Auriga--Cal itself is in an earlier evolutionary stage \citep{BroekhovenFieneetal2014a}. An H-R diagram analysis of the \lkha 101 cluster alone (where it is difficult to measure the infrared class ratios due  to the bright emission around \lkha 101) suggests that the majority of individual YSOs have ages $<$ 3 Myr with a median age of 1 Myr \citep{Wolketal2010}. This situation makes Auriga--Cal an interesting target in which to study both YSOs and cloud properties at early evolutionary stages. \cite{Harveyetal2013} observed Auriga--Cal with PACS (\citealt{PACS}) at 70 and 160~\micron\ and SPIRE (\citealt{SPIRE}) at 250, 350, and 500~\micron\ on the \Herschel\ \textit{Space Observatory} and Bolocam at the Caltech Submillimeter Observatory (CSO) at 1.1 mm  to map the large-scale structure and identify Class 0/I YSOs with \Herschel/PACS and Bolocam photometry. In this work, we focus on the protostellar objects (YSOs) evident in submillimeter observations.

Submillimeter observations probe the cool, optically thin thermal emission from the dust of YSOs and their nascent clouds. This makes such wavelengths optimal for measuring dust masses, as observations of the YSOs probe the cool material of the circumstellar envelope and the disk. The circumstellar envelope (expected to have sizes up to $\sim$10 000 au, $\sim$22\arcsec\ at Auriga--Cal's distance) is present in the earliest stages of star formation and dissipates as material is transferred onto the young star through the disk (expected to have sizes up to $\sim$100 au, $\sim$0.2\arcsec\ at Auriga--Cal's distance). The YSOs are identifiable by their compact emission in comparison to the more diffuse cloud.

We present the first results from observations of Auriga--Cal taken with the Submillimetre Common-User Bolometer Array-2 (SCUBA-2; \citealt{Hollandetal2013}) on the James Clerk Maxwell Telescope (JCMT). These data are part of the JCMT Gould Belt Legacy Survey (GBS; \citealt{WardThompsonetal2007}) to observe nearby (within 500 pc) star-forming regions and trace the earliest stages of star formation. We also include previously unpublished \co\ $J=3-2$ observations (PI: Matthews; program IDs M09BC16 and M10BC09) taken with the Heterodyne Array Receiver Programme (HARP). In this work, we describe the observations and data reduction in Section~\ref{sec:obs}. In Section~\ref{sec:sources}, we describe the source extraction (Section~\ref{sec:getsources}) to identify compact sources associated with protostellar objects and isolate them from larger structures such as cloud emission and clumps. We highlight the locations of these candidate YSOs within the cloud in Section~\ref{sec:locations}. We compare our candidate YSO catalog with the \Spitzer\ and \Herschel/PACS YSO catalogs (Section~\ref{sec:previouscatalogs}) to identify robust YSOs and previously unknown young objects. We also describe the measurement of fluxes (Section~\ref{sec:fluxes}) and measure the limit on possible contamination of our 850~\micron\ fluxes with CO emission (Section~\ref{sec:source_CO}). We use the submillimeter emission to measure the circumstellar masses of YSOs (Section~\ref{sec:masses}). Finally, we compare the population of embedded candidate YSOs in Auriga--Cal to that in Orion A to investigate the recent relative star formation rates between the two clouds (Section~\ref{sec:Orion}). We summarize our conclusions in Section~\ref{sec:summary}.


\section{Observations and Data Reduction}
\label{sec:obs}

\subsection{SCUBA-2}
\label{sec:scuba2}

\begin{figure*}[h]
\includegraphics[trim=0cm 0cm 0cm 0cm, clip=True, width=7.0in]{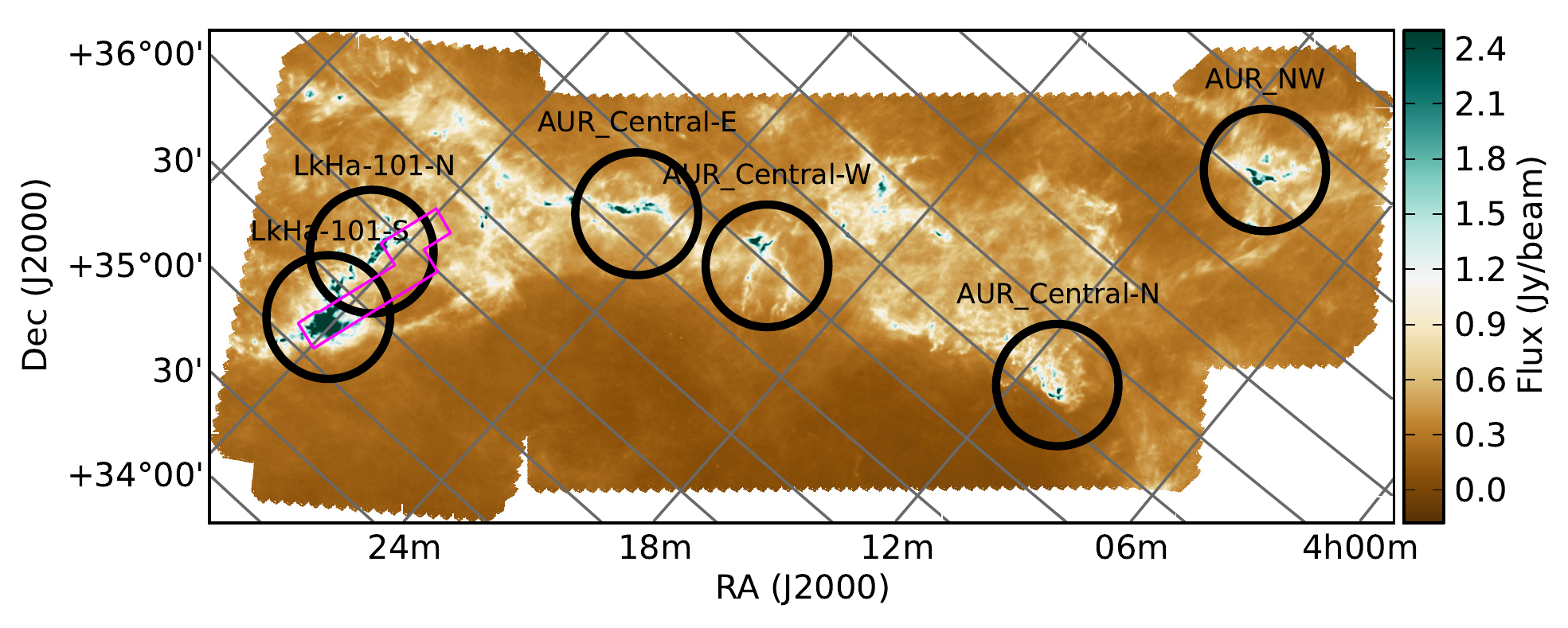}\centering
\caption{\footnotesize{SCUBA-2 observed regions in Auriga--Cal. Circles marking the area observed by SCUBA-2 (and labeled according to their observation name) are overlaid on the \Herschel\ 500~\micron\ map from \citealt{Harveyetal2013} to illustrate the locations throughout the cloud that were mapped. The regions with the highest column density and most compact sizes were targeted, and the \lkha 101 pongs, which cover the part of the cloud with the densest area of star formation, were prioritized to be observed in the best weather (Band 1). The \co\ $J=3-2$ coverage is outlined in magenta. For optimal display of the entire cloud, the celestial coordinates are tilted; i.e., north is not up as it is in Figure~\ref{fig:pongs}.}\label{fig:footprint}}
\vspace{-3mm}
\end{figure*}

\begin{figure*}
\includegraphics[trim=0cm 4cm 0cm 0.5cm, clip=True, width=6.5in,page=1]{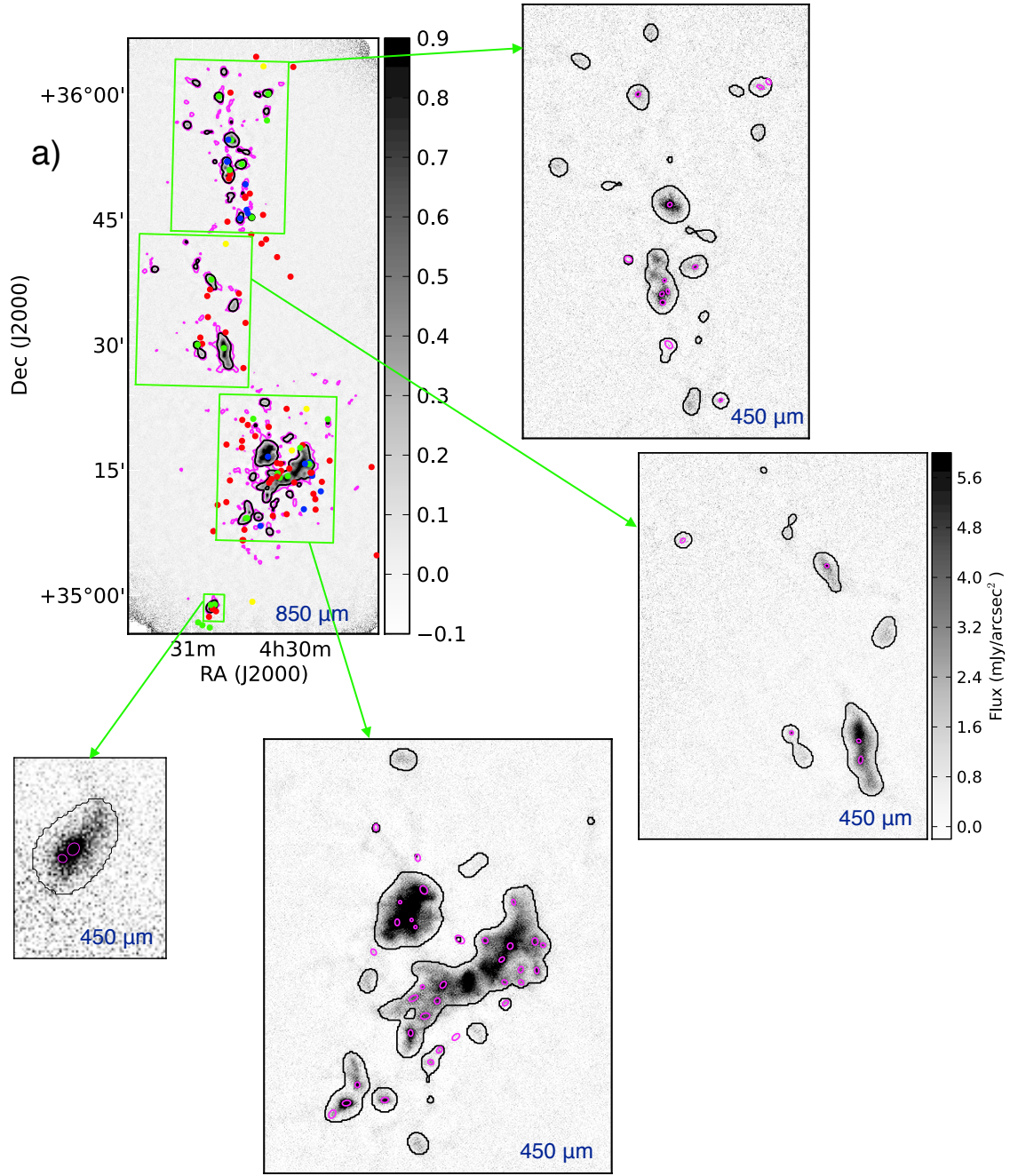}\centering
\caption{\footnotesize{(a) Maps of the \lkha 101 regions at 850~\micron\ (top left) and 450~\micron\ (other panels). The continuation of this figure on following pages shows the maps of the AUR\_Central-E (b), AUR\_Central-W (c), AUR\_Central-N (d), and AUR\_NW (e) regions at 850~\micron\ (left) and 450~\micron\ (right). Black contours in all panels trace the external mask used in the data reduction (Section~\ref{sec:scuba2}). In the 850~\micron\ panel, magenta contours highlight detected emission tracing an S/N of 1 smoothed over 5 pixels (computed from the data and variance maps). Overlaid are green, blue, red, and yellow points that show the locations of Class I, flat-spectrum, Class II, and Class III YSOs, respectively, from \cite{BroekhovenFieneetal2014a}. (See discussion in Section~\ref{sec:locations}.) The green boxes in the 850~\micron\ panel show the areas that are enlarged for the 450~\micron\ panels. The 450~\micron\ panels are all displayed with the same color-scale ranges. In these panels, magenta ellipses mark identified candidate YSOs in SCUBA-2 maps, with major and minor FWHM and orientation according to the source properties measured with \getsources\ (see Section~\ref{sec:getsources} and Figures~\ref{fig:vet} and \ref{fig:appendix} for panels of individual candidate YSOs). Some faint filamentary structure is visible in these maps and matches that seen in \Herschel\ 500~\micron\ maps (c.f. Figure~\ref{fig:footprint}). (Note that some of the noisy map edges are visible in the area displayed.) 
}\label{fig:pongs}}
\vspace{-3mm}
\end{figure*}

\begin{figure*}
\ContinuedFloat
\includegraphics[trim=2.9cm 7cm 2.3cm 6cm, clip=True, width=6.5in,page=2]{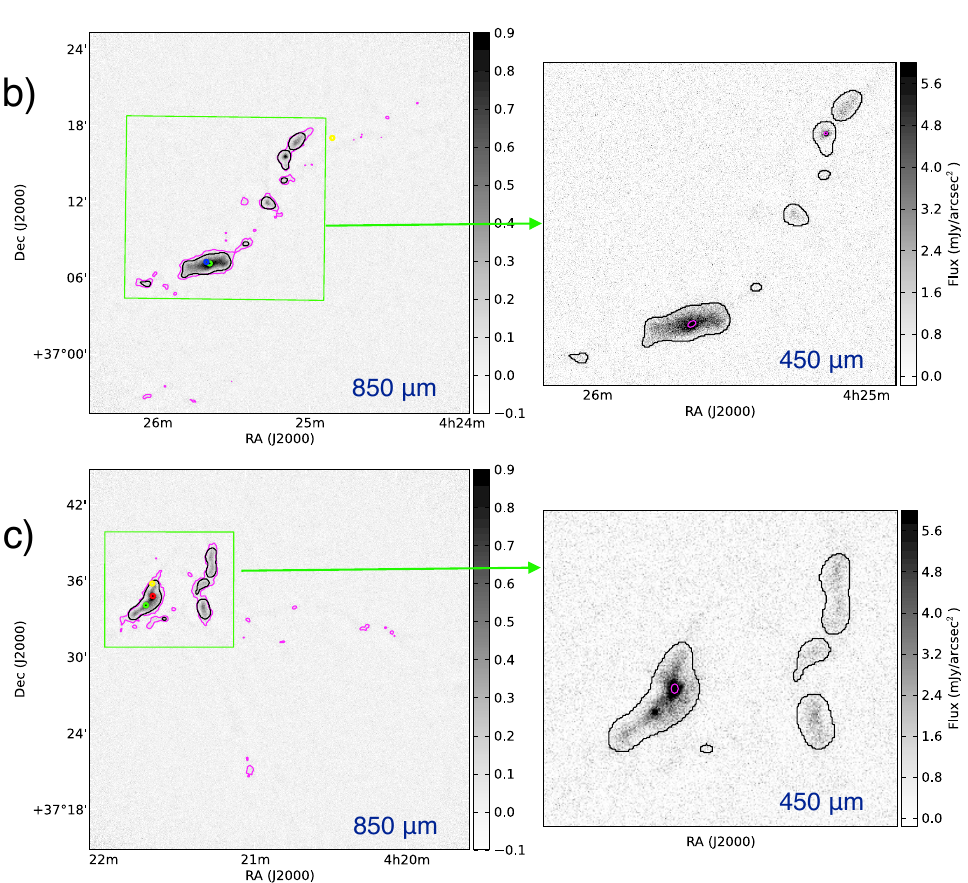}\centering\\
\caption{\small{\textit{continued}. }}
\vspace{-3mm}
\end{figure*}

\begin{figure*}
\ContinuedFloat
\includegraphics[trim=2.5cm 6cm 2.7cm 6cm, clip=True, width=6.5in,page=3]{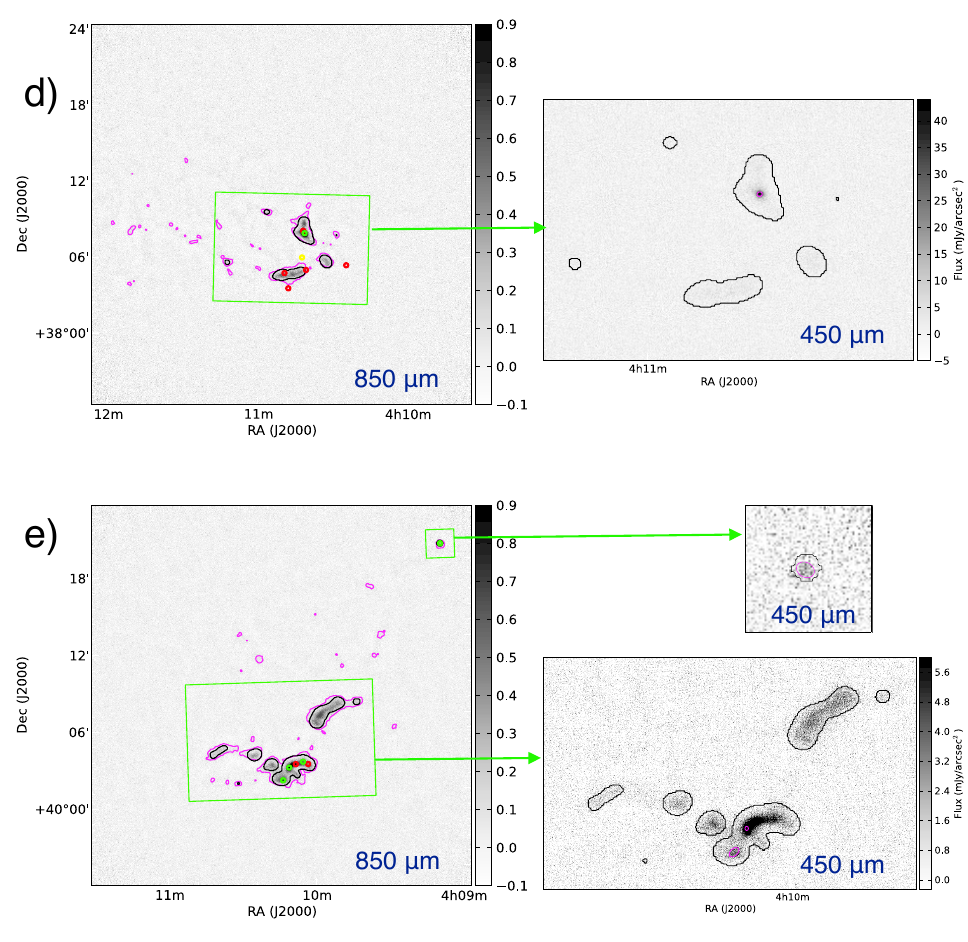}\centering\\
\caption{\small{\textit{continued}. }}
\vspace{-3mm}
\end{figure*}

Continuum observations at 850 and 450~$\mu$m were made using fully sampled 30\arcmin\ diameter circular regions, referred to as ``pongs'' (PONG1800 mapping mode; \citealt{Kackleyetal2010}) between 2012 July and 2015 January.  Larger regions were mosaicked with overlapping scans. The reduced data presented here are from the GBS Legacy Release 1 of the GBS data reduction team \citep{Mairsetal2015}. Six different pong regions were observed, as shown in Figure~\ref{fig:footprint}. Only the dense areas of the cloud were observed as part of the larger goal of the GBS to cover as many regions of \Av\ $\gtrsim$ 3 as possible within the finite allocation given to the survey (roughly two-thirds of the cloud above this extinction level). For wispier clouds, such as Auriga--Cal, this results in more piecemeal coverage as compared to Orion A, for example. The pong regions observed were chosen on the basis of having the highest column densities and most compact sizes in the \Herschel\ data. The AUR\_Central-N region was added to the survey in 2015 January when the management of the JCMT by the Joint Astronomy Centre was coming to a close and the legacy surveys were nearing completion. Extra regions that could be observed in Band 2 weather were submitted so the JCMT would not be idle were there no other higher-priority legacy survey regions visible. We submitted the AUR\_Central-N region as it contained one of the few groups of YSOs identified with \Spitzer\ \citep{BroekhovenFieneetal2014a} not already included in the survey coverage. Regions within the GBS were prioritized such that the highest-priority regions were observed in Band 1 ($\tau_{\rm 225 GHz} < 0.05$), the best weather conditions, to have better 450~\micron\ sensitivity, with other regions observed in Band 2 weather ($0.05 < \tau_{\rm 225 GHz} < 0.08$). The \lkha 101 pongs in the southern end of the cloud were prioritized for Band 1 observations. As this region of the cloud has higher-density material, it unsurprisingly is the richest area in the cloud in terms of previously identified YSOs. Band 1 regions are observed with four repeats each, and regions observed in Band 2 weather have six repeats each (except AUR\_Central-N, which has 5 repeats). The 450 and 850~\micron\ maps of each region are shown in Figure \ref{fig:pongs}.

The data were reduced using an iterative mapmaking technique (\textit{makemap} in {\sc smurf}; \citealt{Chapinetal2013}) and gridded to 3\arcsec\ pixels at 850~$\mu$m and 2\arcsec\ pixels at 450~$\mu$m. The iterations were halted when the map pixels, on average, changed by $<$0.1\% of the estimated map rms. The initial reductions of each individual scan were coadded to form a mosaic from which a signal-to-noise ratio (S/N) mask of S/N $>$ 3 was produced for each region at 850~\micron, which was then smoothed and rethresholded in an attempt to bridge nearby areas of bright emission likely containing emission. This better determines the locations of the fainter emission, as the coadded mosaic of multiple pongs has a higher S/N than the individual pongs. The final mosaic was produced from a second reduction using this mask for both 850 and 450~\micron\ maps to define areas of emission. As discussed in \cite{Mairsetal2015}, detection of emission structure and calibration accuracy are robust within the masked regions and uncertain outside of the masked regions. Any astronomical signal that may be outside the mask, although real, may likely be underestimated in flux and size. The mask used in the reduction can be seen in the quality array in the reduced data file and is shown in Figure~\ref{fig:pongs}.

A spatial filter of 600\arcsec\ is applied to the individual time series in the data reduction, which means that flux recovery is robust for sources within the masked region and  with a Gaussian FWHM less than 2.5\arcmin. This filter is applied to prevent the growth of large, unreal structures in the final maps. Sources between 2.5\arcmin\ and 7.5\arcmin\ will be detected, but both the flux and the size are underestimated because Fourier components with scales greater than 5\arcmin\ are removed by the filtering process. Detection of sources larger than 7.5\arcmin\ is dependent on the mask used for reduction (see \citealt{Mairsetal2015} for more details).

The data are calibrated in mJy~arcsec$^{-2}$, using aperture flux conversion factors of  2.34 and  4.71~Jy/pW/arcsec$^{2}$ at 850 and 450~$\mu$m for absolute flux calibration, respectively, derived from average values of JCMT calibrators \citep{Dempseyetal2013}, and correcting for the pixel area. The pong scan pattern leads to lower noise in the map center and overlap regions, while data reduction and emission artifacts can lead to small variations in the noise over the whole map.

The typical pixel-to-pixel noise level in the 850~\micron\ maps is 0.05 mJy~arcsec$^{-2}$. The noise level varies more for 450~\micron, which is more sensitive to the different conditions in which the data were taken (for example, weather conditions and extended, i.e., daytime, observing), but is typically 1~mJy~arcsec$^{-2}$. It is twice that for AUR\_Central-N (2.2~mJy~arcsec$^{-2}$, observed in weather fluctuating between Band 2 and Band 3 conditions) and slightly lower (0.7~mJy~arcsec$^{-2}$) for LkHa-101-S (not taken during extended observing like LkHa-101-N was). The detected emission (Figure~\ref{fig:pongs}) shows filamentary structure reminiscent of the large-scale structure observed with \Herschel/SPIRE (\citealt{Harveyetal2013}; see, for example, Figure~\ref{fig:footprint}). There are some locations in the map, particularly near \lkha 101, with negative bowling around bright emission. This artifact occurs when the boundary of the external mask, which forces the flux to go to zero at the edge where it meets the noise level, does not contain all of the true emission. Any future work on the larger-scale cloud emission will need a more appropriate mask to recover such emission. The mask used in this work, however, is sufficient for recovering compact sources. Reductions testing different external masks for the JCMT GBS showed that the flux of a compact source, measured with aperture photometry, is consistent between reductions, as the increase in recovered large-scale emission is accounted for with the sky aperture. We therefore continue our analysis, which is focused on the compact sources in Auriga--Cal associated with YSOs, with the standard external mask described above.

All maps and data products associated with this paper are available at https://doi.org/10.11570/17.0008.
More recent improved reductions may be publicly available from the GBS.


\subsection{HARP}

We include previously unpublished \co\ $J=3-2$ (hereafter CO) observations (PI: Matthews; program IDs M09BC16 and M10BC09) taken with HARP. Although that program was not completed, the coverage around \lkha 101, the region most susceptible to CO contamination (see below), was completed by the GBS with the same observing setup as the PI data. The area observed with HARP is shown in Figure~\ref{fig:footprint}.


All HARP data were processed with the ORAC-DR heterodyne pipeline \citep{Jennessetal2015} using the \textsc{reduced\_science\_narrowline} recipe.  In brief, this sorts the time series into temporal order and identifies and rejects spectra affected by high-frequency noise and low-frequency non-astronomical signal using a non-linearity coefficient of 0.08 (where the best spectra have coefficients $<$0.025). The recipe enters an iterative phase.  First, it combines all the filtered time series cubes to form a group spectral cube with 6\arcsec\ pixels and an effective spatial resolution of 16.6\arcsec, 1.0\kms\ (\lkha~101) or 0.1\kms\ spectral resolution.  The spectral cube is smoothed with a spatial bias, and linear baselines are subtracted to enable emission features to be detected and masked.  The emission-free regions permit improved baseline fits, which are then subtracted from the group cube.  One iteration proved sufficient.  A clump-finding algorithm applied to the group cube locates the emission (using the \textsc{Clumpfind} technique from the CUPID package; \citealt{CUPID}), which is integrated to generate a map for the CO contamination.

\subsubsection{CO decontamination}
\label{sec:codecontamination}

The \co\ $J=3-2$ emission line lies within the 850~\micron\ SCUBA-2 filter, and therefore such emission is included in the total flux observed at 850~\micron. We use the HARP observations to remove the CO contribution from the 850~\micron\ maps in order to isolate the dust continuum emission. The CO emission has been found to be a significant contaminant of observations of the dust continuum in the presence of outflows from young YSOs \citep{Drabeketal2012}; \cite{Sadavoy2013} found that the CO line emission contributed up to 90\% of the 850~\micron\ flux in the presence of outflows. It is therefore necessary to measure the CO flux in the NGC 1529 cluster area around \lkha 101 where we expect the highest contamination, as it hosts the brightest cloud emission and is the densest area of star formation in the cloud \citep{BroekhovenFieneetal2014a}. We subtract the detected CO emission in this one region to place an upper limit on CO contamination elsewhere.

To create an 850~\micron\ map that is decontaminated of CO emission, the 850~\micron\ data are reduced in the same way as the external mask reduction described in Section~\ref{sec:scuba2}, with the exception of supplying the integrated CO intensity map as a negative source to the \textsc{makemap} routine. Done in this way, the CO emission that is subtracted from the map is subject to the same processing effects (such as spatial filtering) as the 850~\micron\ data are. The CO contamination at YSO locations is discussed in Section~\ref{sec:source_CO}.


\section{Results}
\label{sec:sources}

\subsection{Identifying Candidate YSOs}
\label{sec:getsources}

As described in Section~\ref{sec:scuba2}, the emission in the SCUBA-2 maps is composed of large-scale cloud emission and compact emission from YSOs. It is nontrivial to isolate the large-scale cloud emission from the compact emission associated with YSOs (expected to have sizes up to $\sim$10 000 au, $\sim$22\arcsec\ at Auriga--Cal's distance), especially with the large beam sizes of single-dish submillimeter observatories such as the JCMT (14.5\arcsec\ and 7.5\arcsec\ at 850 and 450~\micron, respectively). There are many source-finding algorithms used for such datasets (i.e., single-dish submillimeter observations of star-forming regions), each with its own technique for identifying and characterizing emission structure. We use the \getsources\ algorithm (\citealt{Menshchikovetal2012}; version 1.140127) to identify sources due to its sophisticated approach of using spatial decompositions and handling information from multiple maps with different resolutions. These qualities are especially powerful for our multiwavelength maps of varying resolution at 850 and 450~\micron, especially when complementing with information from \Herschel/SPIRE maps. This approach allows us to retain the advantage of the highest resolution available with JCMT maps, rather than having to degrade the resolution of the 450~\micron\ maps to match that of the 850~\micron\ maps (or to degrade the SCUBA-2 maps to the resolution of the \Herschel/SPIRE maps). This is particularly important  for the southern end of the cloud, where the star formation density is highest (and therefore source crowding is more of an issue), and the region around \lkha 101, where compact identification is further complicated by the bright cloud emission warmed by the early-B star. We start by identifying all sources and then continue our analysis with only those that are compact, and therefore likely associated with YSOs (as opposed to larger sources associated with clumps and starless cores) in order to identify the population of submillimeter protostars and measure the mass of their circumstellar material. 

The \getsources\ algorithm was developed for source extraction in the \Herschel\ Gould Belt Survey \citep{Andreetal2010}. It identifies sources by decomposing the maps into different spatial scales and using multiwavelength observations of fields to identify structures and sources common to different maps while accounting for various resolutions. An initial extraction is run at each wavelength independently (monochromatic extractions), and then a combined extraction is done using information from the monochromatic extractions to make a source catalog. A final extraction (also composed of  first monochromatic extractions and then a combined extraction) then uses the combined catalog from the initial extraction to flatten the images by better modeling the background cloud emission and measures the source properties from these flattened maps.

We perform a separate source extraction for each field (first cropped to exclude the noisy edges) observed by SCUBA-2 independently. This is because of the varying noise levels between the 450~\micron\ maps of different fields due to the increased sensitivity to the weather conditions in which they were observed (Section~\ref{sec:scuba2}).
The exception is the pong regions LkHa-101-N and LkHa-101-S, which overlap, and therefore the extraction is performed on a mosaic of these regions. This allows us to identify sources in the overlap region that are at the noisy edges of the individual pongs and therefore would otherwise be excluded. The observations of these two regions are similar because they were both observed in Band 1 weather. 

We also take advantage of the \Herschel/SPIRE maps from \cite{Harveyetal2013}. The \Herschel/SPIRE maps are first processed with the SCUBA-2 mapmaker so that maps from both instruments are spatially filtered in a similar way. (Both \Herschel/SPIRE and SCUBA-2 maps are subject to spatial filtering in their mapmaking processes; however, SCUBA-2 maps have much more large-scale structure filtered out due to the nature of filtering out the atmosphere with ground-based submillimeter observations.) Processing the \Herschel/SPIRE maps is described in detail in \cite{Chenetal2016}. Briefly, the \Herschel/SPIRE maps are included in the reduction of SCUBA-2 data as a positive source, albeit as a small fluctuation with respect to the SCUBA-2 emission, by first scaling the \Herschel/SPIRE maps by an arbitrary constant, $c$. (This is similar to the process to remove CO emission from the 850~\micron\ maps, described in Section~\ref{sec:codecontamination}, except that the \Herschel/SPIRE maps are included as a positive source rather than the integrated CO map that was included as a negative source.) The original SCUBA-2-only map is then subtracted from this SCUBA-2 + $c$\Herschel/SPIRE map to isolate the filtered \Herschel/SPIRE emission. The resulting map is then unscaled by the arbitrary constant to recover the actual level of emission. Processing \Herschel/SPIRE maps in this way to include them in analysis of SCUBA-2 maps has proved to be advantageous when measuring the properties of clumps and cloud emission across star-forming regions, as shown in \cite{Sadavoy2013}, \cite{Chenetal2016}, and \cite{WardThompsonetal2016}. Ward-Thompson et al. showed that  the 250~\micron\ SPIRE maps filtered in this way sample the same material probed by the 850~\micron\ SCUBA-2 maps. This is because the warmer, largest-scale cloud emission is filtered out from  the cooler cloud clumps. Sadavoy et al. and Chen et al. showed that this technique is necessary for measuring temperature and $\beta$ variations in Perseus.

We include the resulting \Herschel/SPIRE maps processed with the SCUBA-2 mapmaker as measurement-only images in the \getsources\ extraction in order to include fluxes of SCUBA-2 sources also measured at 250, 350, or 500~\micron. (We do not run \getsources\ on the \Herschel/PACS maps, however, as source identification in these maps was already done by \citealt{Harveyetal2013}.) This means that \getsources\ uses all the maps (SCUBA-2 and \Herschel/SPIRE) to model the large-scale structure to better isolate it from smaller-scale sources. This results in better modeling overall of the sources in the SCUBA-2 maps without attempting to identify and characterize all sources in the \Herschel/SPIRE maps (which is beyond the scope of this work). 

The final \getsources\ catalog contains 223 sources in SCUBA-2 maps and the extracted fluxes and sizes of the sources at each wavelength, as well as various internal parameters to represent the quality or robustness of each extracted source. This initial source catalog contains various kinds of sources that can be appear as a 2D Gaussian structure in these maps, such as large-scale cloud emission, clumps, cores, and YSOs/protostars. Our analysis is targeted only at the YSO/protostar population, which we expect to have sizes up to $\sim$10 000 au, $\sim$22\arcsec\ at Auriga--Cal's distance and $\sim$26\arcsec\ and $\sim$23\arcsec\ when convolved to the 850 and 450~\micron\ beams. Therefore, we first select only the compact sources within the \getsources\ extraction catalog of 223 submillimeter sources and then visually confirm this subset. The cuts for compact sources associated with protostars  are based on geometry (sources must be compact with FWHM $\leq 30$\arcsec\ along both major and minor axes and must not be elongated, i.e., aspect ratio $\leq 2$) and flux (having a positive flux value with an S/N $\geq 3$ from \getsources's internal parameters). 

\begin{figure*}
\includegraphics[trim=3.5cm 0.cm 3.8cm 0cm, clip=True, width=6.5in]{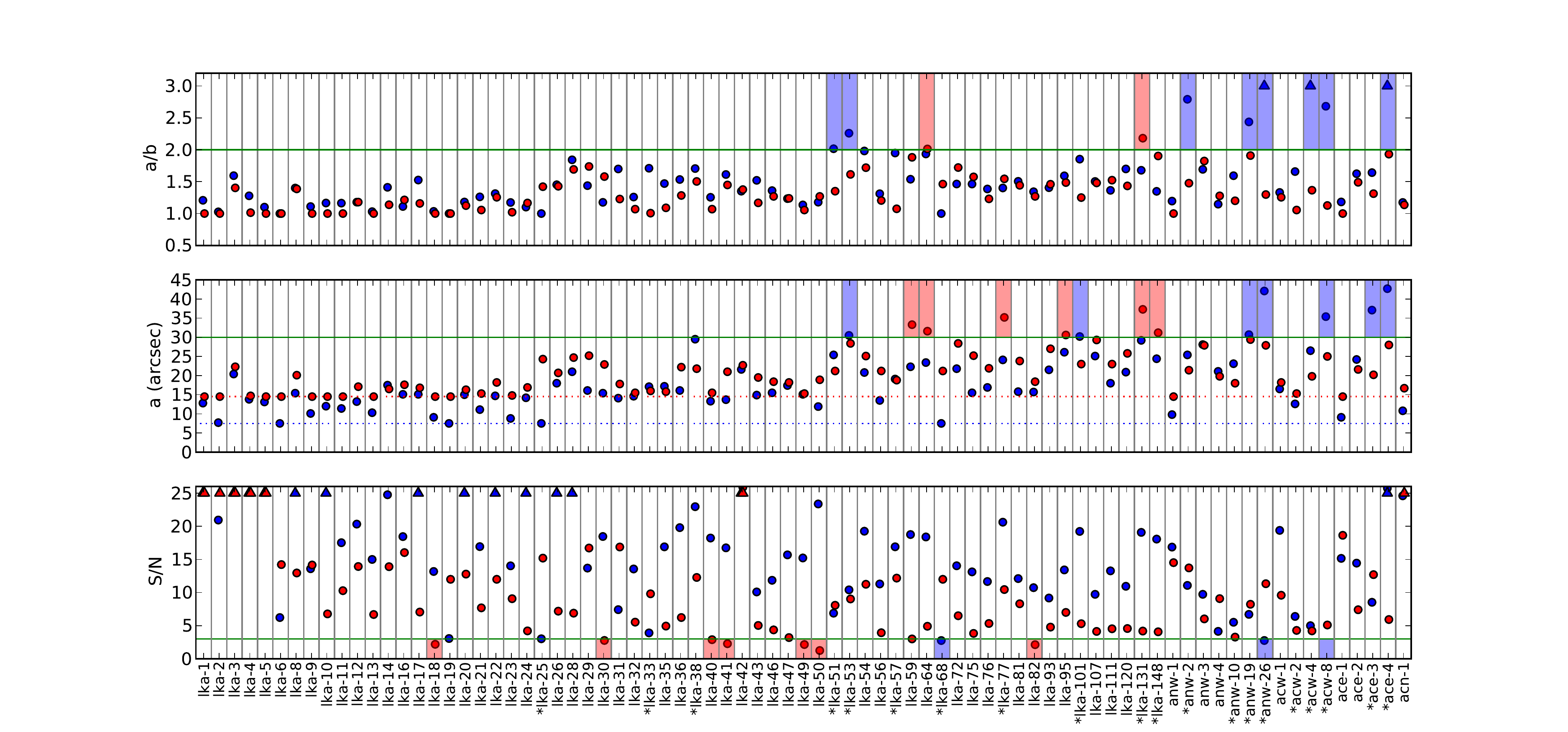}\centering \\ 
\caption{\footnotesize Criteria for identifying compact objects from the \getsources\ extraction: aspect ratio (top), size (middle), and S/N (bottom) for each of the 79 objects that meet our criteria for compact sources at one or both SCUBA-2 wavelengths.  The green horizontal lines mark the boundaries for each criterion during the initial selection of compact sources. The measured property is displayed for each source at both 450~\micron\ (blue) and 850~\micron\ (red). (Note that an upward arrow is displayed for values that extend beyond the plot boundaries.) The area of the plot is shaded where a compact source does not meet the criterion at the wavelength corresponding to the color of the shading (blue=450, red=850). Blue and red dotted lines show the 450 and 850~\micron\ beam sizes, respectively. Additional flags arise during the vetting process if a source is much larger at 450~\micron\ than at 850~\micron\ (since the 850~\micron\ beam is larger and we expect a real protostellar source to have roughly the same physical size at both wavelengths) or when the elongation measured at the two wavelengths is very different. Each compact source is labeled according to the internal \getsources\ ID from the source extraction. Those preceded by a `*' are excluded from the final list of YSO candidates based on the flagging described.\label{fig:criteria}}
\end{figure*}

\begin{figure*}
\includegraphics[trim=2.25cm 3.7cm 0cm 11.cm, clip=True, width=7.5in,page=1]{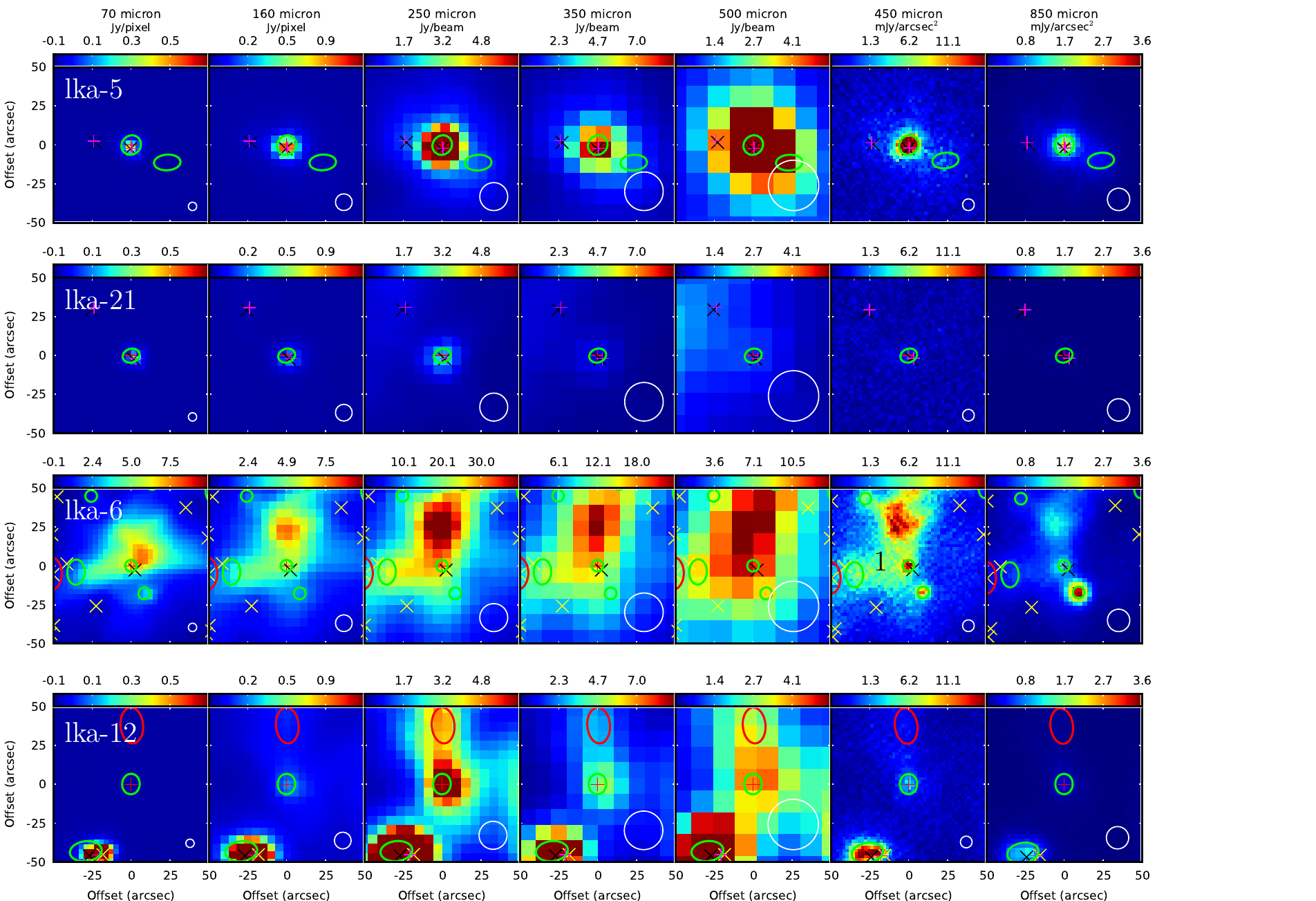}
\caption{{\footnotesize Examples of the quality-assurance maps for compact sources extracted using \getsources\ showing (left to right) the \Herschel/PACS (70 and 160~\micron), \Herschel/SPIRE (250, 350, and 500~\micron), and SCUBA-2 (450 and 850~\micron) maps. Each panel is centered on the compact source in question which is marked with crosshairs. Elliptical regions are the same as in Figures~\ref{fig:pongs} with blue, red, and green colors marking sources that satisfy the compact source criteria at 450~\micron, 850~\micron, or both, respectively, and with major and minor FWHM and orientation according to the source properties measured with \getsources. The internal \getsources\ ID is listed in the upper left corner of the 70~\micron\ map for each source. From top to bottom, the rows show the following. (1) An example of a well-detected compact source (lka-5) associated with a YSO identified with both \Spitzer\ and \Herschel/PACS. (2) A compact source (lka-21) identified in SCUBA-2 maps that is not very convincing visually at either 450 or 850~\micron, but for which inspection of \Herschel\ maps provided the by-eye conviction of the presence of compact emission. This demonstrates the effectiveness of \getsources's source identification in the SCUBA-2 maps. (3) An example of a compact source (lka-6) for which the presence of compact 70~\micron\ emission, indicative of a protostellar source, can be neither confirmed nor ruled out given the presence of bright background emission. (4) An example of a seemingly robust submillimeter compact source (lka-12) that is not detected at wavelengths shorter than 160~\micron. Each of these four compact sources passes the vetting process to be identified as candidate YSOs. Such figures for the remaining compact sources are included in the Appendix and shown in Figure~\ref{fig:appendix}.}\label{fig:vet}}
\end{figure*}

Figure~\ref{fig:criteria} shows the measured values for each of the 79 compact sources that meet these criteria at 450 and/or 850~\micron\ and highlights where a compact source does not meet a specific criterion at either wavelength. Concerns arise during the vetting process if a source is much larger at 450~\micron\ than at 850~\micron\ (since the 850~\micron\ beam is larger and we expect the source to have the same physical size at both wavelengths) or when the elongation measured at the two wavelengths is very different. These flags are considered along with the visual inspection of the sources. We plot the location of each compact source in a zoomed-in region of the SCUBA-2 maps to inspect them more carefully. We similarly also plot their location in \Herschel/SPIRE and \Herschel/PACS maps. These figures help us to determine (1) the reliability of each extracted compact source; (2) whether there is, or could be, compact 70~\micron\ emission indicating a protostellar source; and (3) the shortest wavelength at which the compact source is evident. We show an example of each of these points in Figure~\ref{fig:vet}. (The full collection of plots for each compact source is included in the Appendix.) Following this inspection, 20 compact sources are removed. These are sources that were generally associated with a tail feature from background emission near a very bright compact source or met the criteria at only one wavelength and appeared to be (faint) extended cloud emission. We refer to the remaining 59 vetted compact sources identified with the \getsources\ algorithm as candidate YSOs, due to the compact nature of their emission. They are listed in Table~\ref{tbl:getsources_sources} and named according to the IAU convention and designation for the GBS.

\begin{deluxetable}{lcccccccccl}
\rotate
\tabletypesize{\tiny}
\tablewidth{0pc}
\tablecaption{Possible YSOs identified in SCUBA-2 maps with \getsources
\label{tbl:getsources_sources}}
\tablehead{ 
\ch{ID} &         \ch{Source Name} &     \ch{Internal \getsources} & \mc{2}{c}{IR YSO Catalog Identifiers} & \ch{450~\micron\ Flux} & \ch{850~\micron\ Flux} & \ch{250~\micron\ Flux} & \ch{350~\micron\ Flux} & \ch{500~\micron\ Flux} &    \ch{Overestimated} \\ 
  \ch{} &                    \ch{} & \ch{Source Catalog ID\tnm{a}} &      \ch{\Spitzer} &     \ch{\Herschel} &              \ch{(Jy)} &              \ch{(Jy)} &              \ch{(Jy)} &              \ch{(Jy)} &             \ch{(Jy)}  & \ch{Flux Notes\tnm{b}} } 
\startdata 
      1 & JCMTLSG J043048.1+345841 &                         lka-1 &           \nodata &           \nodata &       10.30 $\pm$ 1.08 &        2.43 $\pm$ 0.26 &       15.02 $\pm$ 0.80 &        9.06 $\pm$ 0.47 &        5.61 $\pm$ 0.30 &                   IR,N \\  
      2 & JCMTLSG J043014.7+351625 &                         lka-2 &           \nodata &           \nodata &        8.09 $\pm$ 0.91 &        2.72 $\pm$ 0.39 &       11.43 $\pm$ 1.82 &        8.25 $\pm$ 0.87 &        8.32 $\pm$ 0.48 &                    C,N \\  
      3 & JCMTLSG J043028.3+350919 &                         lka-3 &                88 &           \nodata &        6.92 $\pm$ 0.72 &        1.11 $\pm$ 0.12 &       30.79 $\pm$ 1.61 &       17.02 $\pm$ 0.91 &        6.92 $\pm$ 0.37 &                     IR \\  
      4 & JCMTLSG J043038.6+355025 &                         lka-4 &           \nodata &                42 &        6.99 $\pm$ 0.74 &        1.37 $\pm$ 0.18 &        7.76 $\pm$ 0.44 &        5.98 $\pm$ 0.35 &        2.88 $\pm$ 0.22 &                      C \\  
      5 & JCMTLSG J043036.8+355439 &                         lka-5 &               103 &                38 &        9.50 $\pm$ 0.99 &        1.70 $\pm$ 0.20 &       24.60 $\pm$ 1.26 &       12.73 $\pm$ 0.68 &        7.83 $\pm$ 0.43 &                      N \\  
      6 & JCMTLSG J043015.4+351642 &                         lka-6 &               138 &           \nodata &       10.48 $\pm$ 1.16 &        1.81 $\pm$ 0.32 &       47.98 $\pm$ 2.93 &       17.59 $\pm$ 1.19 &        4.78 $\pm$ 0.32 &                    C,N \\  
      7 & JCMTLSG J043015.6+351209 &                         lka-8 &           \nodata &           \nodata &        5.47 $\pm$ 0.58 &        0.72 $\pm$ 0.12 &        6.96 $\pm$ 0.74 &        5.49 $\pm$ 0.60 &        4.24 $\pm$ 0.33 &                      C \\  
      8 & JCMTLSG J043038.0+355103 &                         lka-9 &               107 &                40 &        4.16 $\pm$ 0.48 &        0.75 $\pm$ 0.16 &        4.89 $\pm$ 0.29 &        3.58 $\pm$ 0.25 &        2.12 $\pm$ 0.19 &                      C \\  
      9 & JCMTLSG J043041.4+352943 &                        lka-10 &               117 &                45 &        6.21 $\pm$ 0.67 &        1.19 $\pm$ 0.18 &        6.16 $\pm$ 0.73 &        6.16 $\pm$ 0.44 &        5.51 $\pm$ 0.33 &                        \\  
     10 & JCMTLSG J043048.7+353755 &                        lka-11 &               124 &                50 &        4.01 $\pm$ 0.45 &        0.74 $\pm$ 0.14 &        5.22 $\pm$ 0.34 &        3.95 $\pm$ 0.24 &        2.35 $\pm$ 0.19 &                        \\  
     11 & JCMTLSG J043026.1+351003 &                        lka-12 &           \nodata &           \nodata &        3.60 $\pm$ 0.40 &        0.54 $\pm$ 0.11 &        5.94 $\pm$ 0.51 &        6.10 $\pm$ 0.42 &        2.81 $\pm$ 0.21 &                        \\  
     12 & JCMTLSG J043056.8+353006 &                        lka-13 &               135 &                57 &        1.97 $\pm$ 0.25 &        0.33 $\pm$ 0.10 &        3.19 $\pm$ 0.24 &        2.15 $\pm$ 0.18 &        1.26 $\pm$ 0.12 &                        \\  
     13 & JCMTLSG J043020.7+350927 &                        lka-14 &           \nodata &           \nodata &        2.75 $\pm$ 0.30 &        0.43 $\pm$ 0.08 &        4.08 $\pm$ 0.47 &        3.78 $\pm$ 0.32 &        2.30 $\pm$ 0.20 &                        \\  
     14 & JCMTLSG J043038.4+355000 &                        lka-16 &               108 &                41 &        3.59 $\pm$ 0.39 &        0.62 $\pm$ 0.10 &        7.49 $\pm$ 0.42 &        5.33 $\pm$ 0.32 &        2.85 $\pm$ 0.20 &                      C \\  
     15 & JCMTLSG J042950.7+351440 &                        lka-17 &           \nodata &           \nodata &        3.23 $\pm$ 0.36 &        0.53 $\pm$ 0.10 &        3.82 $\pm$ 0.69 &        3.55 $\pm$ 0.64 &        2.40 $\pm$ 0.40 &                   C,IR \\  
     16 & JCMTLSG J043013.3+351401 &                        lka-18 &                67 &           \nodata &        2.94 $\pm$ 0.37 &        0.53 $\pm$ 0.14 &                $<$2.22 &                $<$2.37 &        1.26 $\pm$ 0.28 &                      C \\  
     17 & JCMTLSG J043017.7+351725 &                        lka-19 &           \nodata &           \nodata &        7.84 $\pm$ 0.89 &        1.21 $\pm$ 0.26 &       18.20 $\pm$ 2.02 &       11.21 $\pm$ 0.78 &        8.20 $\pm$ 0.45 &                      C \\  
     18 & JCMTLSG J043044.3+355953 &                        lka-20 &               118 &                46 &        2.55 $\pm$ 0.29 &        0.51 $\pm$ 0.09 &        3.45 $\pm$ 0.22 &        3.95 $\pm$ 0.26 &        2.28 $\pm$ 0.17 &                        \\  
     19 & JCMTLSG J043024.8+354523 &                        lka-21 &                81 &                29 &        1.72 $\pm$ 0.23 &        0.27 $\pm$ 0.09 &        2.84 $\pm$ 0.20 &        1.47 $\pm$ 0.19 &        0.41 $\pm$ 0.12 &                        \\  
     20 & JCMTLSG J043030.8+355141 &                        lka-22 &               100 &                35 &        2.45 $\pm$ 0.28 &        0.54 $\pm$ 0.10 &        1.46 $\pm$ 0.15 &        1.83 $\pm$ 0.18 &        1.95 $\pm$ 0.18 &                        \\  
     21 & JCMTLSG J043049.0+345832 &                        lka-23 &           \nodata &           \nodata &        6.88 $\pm$ 0.78 &        1.43 $\pm$ 0.26 &        3.38 $\pm$ 0.28 &        3.04 $\pm$ 0.20 &        0.40 $\pm$ 0.11 &                      N \\  
     22 & JCMTLSG J043000.9+351553 &                        lka-24 &           \nodata &           \nodata &        3.51 $\pm$ 0.38 &        0.52 $\pm$ 0.09 &                $<$4.80 &                $<$2.80 &        1.53 $\pm$ 0.32 &                      C \\  
     23 & JCMTLSG J043009.2+351406 &                        lka-26 &           \nodata &           \nodata &        5.35 $\pm$ 0.56 &        0.82 $\pm$ 0.11 &        7.53 $\pm$ 1.47 &        7.15 $\pm$ 1.23 &        2.93 $\pm$ 0.30 &                   C,IR \\  
     24 & JCMTLSG J043040.9+352850 &                        lka-28 &           \nodata &           \nodata &        3.94 $\pm$ 0.42 &        0.75 $\pm$ 0.10 &        2.34 $\pm$ 0.46 &       15.55 $\pm$ 0.87 &        8.97 $\pm$ 0.49 &                      C \\  
     25 & JCMTLSG J043018.3+351636 &                        lka-29 &           \nodata &           \nodata &        8.34 $\pm$ 0.87 &        1.22 $\pm$ 0.15 &       37.20 $\pm$ 2.77 &       13.23 $\pm$ 1.14 &        7.20 $\pm$ 0.42 &                 C,IR,N \\  
     26 & JCMTLSG J042955.9+351539 &                        lka-30 &           \nodata &           \nodata &        6.54 $\pm$ 0.68 &        1.03 $\pm$ 0.13 &        6.28 $\pm$ 1.49 &        5.20 $\pm$ 1.30 &        3.71 $\pm$ 0.35 &                      C \\  
     27 & JCMTLSG J043037.2+355032 &                        lka-31 &               106 &                39 &        4.32 $\pm$ 0.48 &        0.84 $\pm$ 0.15 &        4.89 $\pm$ 0.28 &        3.39 $\pm$ 0.21 &        1.99 $\pm$ 0.18 &                      C \\  
     28 & JCMTLSG J043011.6+351058 &                        lka-32 &           \nodata &           \nodata &        1.40 $\pm$ 0.17 &        0.18 $\pm$ 0.06 &        2.18 $\pm$ 0.45 &        1.90 $\pm$ 0.32 &        0.87 $\pm$ 0.18 &                      C \\  
     29 & JCMTLSG J043009.9+351128 &                        lka-35 &           \nodata &           \nodata &        1.06 $\pm$ 0.13 &                $<$0.12 &        2.42 $\pm$ 0.47 &                $<$0.48 &        3.04 $\pm$ 0.23 &                      C \\  
     30 & JCMTLSG J042956.9+351321 &                        lka-36 &           \nodata &           \nodata &        1.13 $\pm$ 0.15 &                $<$0.15 &        4.29 $\pm$ 0.65 &                $<$0.35 &        3.04 $\pm$ 0.23 &                      C \\  
     31 & JCMTLSG J042957.0+351412 &                        lka-40 &           \nodata &           \nodata &        3.25 $\pm$ 0.37 &        0.44 $\pm$ 0.10 &                $<$2.78 &                $<$2.02 &        1.66 $\pm$ 0.30 &                      C \\  
     32 & JCMTLSG J042955.2+351725 &                        lka-41 &           \nodata &           \nodata &        2.76 $\pm$ 0.33 &        0.52 $\pm$ 0.12 &        2.54 $\pm$ 0.73 &                $<$0.65 &        1.66 $\pm$ 0.30 &                      C \\  
     33 & JCMTLSG J043013.0+351755 &                        lka-42 &           \nodata &           \nodata &        5.50 $\pm$ 0.57 &        0.82 $\pm$ 0.09 &       19.42 $\pm$ 1.10 &        9.90 $\pm$ 0.61 &        5.32 $\pm$ 0.32 &                      C \\  
     34 & JCMTLSG J042957.6+351506 &                        lka-43 &           \nodata &           \nodata &        5.95 $\pm$ 0.64 &        1.03 $\pm$ 0.15 &        4.88 $\pm$ 1.46 &                $<$3.74 &        2.16 $\pm$ 0.35 &                      C \\  
     35 & JCMTLSG J042953.8+351442 &                        lka-46 &           \nodata &           \nodata &        2.85 $\pm$ 0.32 &        0.49 $\pm$ 0.09 &        3.01 $\pm$ 0.73 &                $<$0.91 &        2.16 $\pm$ 0.35 &                      C \\  
     36 & JCMTLSG J042951.0+351550 &                        lka-47 &                47 &           \nodata &        3.87 $\pm$ 0.41 &        0.66 $\pm$ 0.09 &                $<$2.74 &                $<$2.79 &        0.86 $\pm$ 0.28 &                   C,IR \\  
     37 & JCMTLSG J043010.4+351326 &                        lka-49 &           \nodata &           \nodata &        3.94 $\pm$ 0.42 &        0.54 $\pm$ 0.09 &                $<$4.26 &                $<$2.68 &        3.08 $\pm$ 0.30 &                      C \\  
     38 & JCMTLSG J042949.4+351541 &                        lka-50 &           \nodata &           \nodata &        2.86 $\pm$ 0.33 &        0.47 $\pm$ 0.11 &                $<$1.48 &                $<$0.76 &        3.08 $\pm$ 0.30 &                      C \\  
     39 & JCMTLSG J043015.6+360014 &                        lka-54 &                70 &                27 &        1.19 $\pm$ 0.14 &        0.32 $\pm$ 0.07 &        1.16 $\pm$ 0.13 &        1.34 $\pm$ 0.16 &        1.05 $\pm$ 0.15 &                        \\  
     40 & JCMTLSG J042953.8+351411 &                        lka-56 &           \nodata &           \nodata &        2.47 $\pm$ 0.29 &        0.38 $\pm$ 0.09 &        2.65 $\pm$ 0.65 &                $<$0.64 &        1.05 $\pm$ 0.15 &                      C \\  
     41 & JCMTLSG J043012.7+351250 &                        lka-59 &           \nodata &           \nodata &        3.35 $\pm$ 0.35 &        0.43 $\pm$ 0.07 &                $<$4.79 &        2.85 $\pm$ 0.91 &        3.42 $\pm$ 0.32 &                      C \\  
     42 & JCMTLSG J043015.1+351333 &                        lka-64 &                68 &           \nodata &        2.91 $\pm$ 0.31 &        0.42 $\pm$ 0.07 &        5.10 $\pm$ 1.25 &        4.25 $\pm$ 0.91 &        2.19 $\pm$ 0.30 &                      C \\  
     43 & JCMTLSG J043031.1+350853 &                        lka-72 &           \nodata &           \nodata &        1.52 $\pm$ 0.17 &        0.21 $\pm$ 0.04 &        1.97 $\pm$ 0.21 &        1.78 $\pm$ 0.19 &        1.36 $\pm$ 0.14 &                        \\  
     44 & JCMTLSG J043022.5+352026 &                        lka-75 &           \nodata &           \nodata &        0.66 $\pm$ 0.10 &                $<$0.13 &        2.29 $\pm$ 0.28 &        1.30 $\pm$ 0.21 &        0.70 $\pm$ 0.14 &                      C \\  
     45 & JCMTLSG J043014.2+351913 &                        lka-76 &           \nodata &           \nodata &        0.46 $\pm$ 0.07 &                $<$0.10 &        2.80 $\pm$ 0.57 &        0.99 $\pm$ 0.25 &        1.11 $\pm$ 0.14 &                      C \\  
     46 & JCMTLSG J043122.0+353905 &                        lka-81 &           \nodata &           \nodata &        1.02 $\pm$ 0.14 &        0.25 $\pm$ 0.07 &        0.46 $\pm$ 0.11 &        1.13 $\pm$ 0.13 &        0.83 $\pm$ 0.11 &                        \\  
     47 & JCMTLSG J043022.9+351525 &                        lka-82 &           \nodata &           \nodata &        0.93 $\pm$ 0.12 &                $<$0.11 &        3.92 $\pm$ 1.23 &        2.41 $\pm$ 0.68 &        1.81 $\pm$ 0.23 &                      C \\  
     48 & JCMTLSG J043046.5+355203 &                        lka-93 &           \nodata &           \nodata &        0.34 $\pm$ 0.05 &        0.08 $\pm$ 0.03 &        0.22 $\pm$ 0.07 &        0.19 $\pm$ 0.05 &        0.13 $\pm$ 0.01 &                        \\  
     49 & JCMTLSG J043037.0+354800 &                        lka-95 &           \nodata &           \nodata &        0.49 $\pm$ 0.06 &        0.15 $\pm$ 0.03 &        0.56 $\pm$ 0.12 &        0.84 $\pm$ 0.15 &        0.87 $\pm$ 0.14 &                        \\  
     50 & JCMTLSG J043005.9+351555 &                       lka-107 &           \nodata &           \nodata &        0.19 $\pm$ 0.03 &                $<$0.05 &        1.64 $\pm$ 0.29 &                $<$0.07 &        0.56 $\pm$ 0.10 &                      C \\  
     51 & JCMTLSG J043013.6+360028 &                       lka-111 &           \nodata &           \nodata &        0.58 $\pm$ 0.08 &        0.18 $\pm$ 0.05 &        0.64 $\pm$ 0.10 &        0.88 $\pm$ 0.13 &        0.57 $\pm$ 0.15 &                        \\  
     52 & JCMTLSG J043006.7+351159 &                       lka-120 &           \nodata &           \nodata &        0.17 $\pm$ 0.03 &                $<$0.03 &        1.08 $\pm$ 0.20 &                $<$0.05 &        0.57 $\pm$ 0.15 &                      C \\  
     53 & JCMTLSG J041008.5+400225 &                         anw-1 &                 6 &                 7 &       11.64 $\pm$ 1.24 &        1.47 $\pm$ 0.24 &       25.38 $\pm$ 1.33 &       12.81 $\pm$ 0.65 &        5.55 $\pm$ 0.30 &                   C,IR \\  
     54 & JCMTLSG J041011.4+400131 &                         anw-3 &                 7 &                 8 &        2.28 $\pm$ 0.24 &        0.47 $\pm$ 0.06 &        3.34 $\pm$ 0.44 &       15.65 $\pm$ 0.85 &        8.19 $\pm$ 0.48 &                        \\  
     55 & JCMTLSG J040902.2+401910 &                         anw-4 &               140 &                 1 &        0.86 $\pm$ 0.11 &        0.16 $\pm$ 0.04 &        2.22 $\pm$ 0.20 &        1.44 $\pm$ 0.15 &        0.79 $\pm$ 0.13 &                        \\  
     56 & JCMTLSG J042138.1+373438 &                         acw-1 &                16 &                12 &        7.26 $\pm$ 0.76 &        0.98 $\pm$ 0.13 &       22.58 $\pm$ 1.16 &       18.77 $\pm$ 0.97 &        8.81 $\pm$ 0.48 &                        \\  
     57 & JCMTLSG J042508.2+371521 &                         ace-1 &           \nodata &                15 &        5.68 $\pm$ 0.65 &        0.77 $\pm$ 0.18 &        5.96 $\pm$ 0.33 &        3.41 $\pm$ 0.22 &        2.10 $\pm$ 0.16 &                        \\  
     58 & JCMTLSG J042538.6+370656 &                         ace-2 &                20 &                16 &        5.95 $\pm$ 0.61 &        0.96 $\pm$ 0.10 &        5.88 $\pm$ 0.55 &        5.06 $\pm$ 0.67 &        9.28 $\pm$ 0.49 &                     IR \\  
     59 & JCMTLSG J041041.2+380754 &                         acn-1 &                10 &                 9 &       23.70 $\pm$ 2.49 &        3.63 $\pm$ 0.47 &       43.04 $\pm$ 2.16 &       21.87 $\pm$ 1.13 &        9.02 $\pm$ 0.51 &                     IR \\  
\enddata
\tablecomments{Uncertainties quoted are statistical flux uncertainties returned by aperture photometry for SCUBA-2 fluxes and \getsources\ for \Herschel/SPIRE fluxes
and include the calibration uncertainties of 10\% and 5\% for SCUBA-2 observations at 450 and 850~\micron\ \citep{Dempseyetal2013}, and $\pm5$ \% for \Herschel/SPIRE fluxes under ideal circumstances (http://herschel.esac.esa.int/hcss-doc-9.0/).}
\tnt{a}{This identifier is to aid comparison with Figures~\ref{fig:criteria}, \ref{fig:vet}, and \ref{fig:appendix} in the Appendix.}
\tnt{b}{We have noted where fluxes are likely overestimated due to bright cloud emission (C), multiple infrared YSOs within the aperture (IR), and/or multiple nearby submillimeter candidate protostars (N).}
\end{deluxetable}

\subsection{Locations of detected candidate YSOs within the cloud}
\label{sec:locations}

As we can see in Figure~\ref{fig:pongs}, the large-scale emission of the cloud is speckled with compact emission from YSOs. The 450~\micron\ panels (Figure~\ref{fig:pongs}, right) show the locations of candidate YSOs (which we identify in Section~\ref{sec:getsources}) against the cloud emission. No candidate YSOs are detected off of the filamentary structure. Such colocation is expected, given that protostars have been observed to lie predominantly along the filaments of their natal clouds \citep{Andreetal2010}. This has two main implications for our sensitivity to YSOs.

Firstly, our sensitivity to YSOs is dependent on their evolutionary stages. As Auriga--Cal is one of the most distant clouds in the GBS survey (and all observations have the same target depth at 850~\micron), we are less sensitive to disk-only YSOs (associated with Class II and Class III YSOs), as opposed to those at earlier stages with a circumstellar envelope as well (associated with Class I and Class F YSOs) and therefore more circumstellar material overall. This is evident when comparing to the 850~\micron\ panels (Figure~\ref{fig:pongs}, left), which show the locations of \Spitzer-identified YSOs from \cite{BroekhovenFieneetal2014a}, color-coded by class. As discussed by \cite{BroekhovenFieneetal2014a}, the Class I and Class F sources, associated with earlier stages of star formation, are found close to the nascent cloud, whereas the Class II and Class III sources, associated with later stages of star formation with less circumstellar material, are more dispersed. The fact that we only detect YSOs in the SCUBA-2 maps that are close to the cloud structure is an immediate reminder that we are sensitive only to the youngest YSOs.

Secondly, since the candidate YSOs are colocated with cloud emission (Figure~\ref{fig:pongs}, right), our sensitivity to protostellar objects is further limited by the brightness of the cloud emission along the line of sight rather than just by our observation sensitivity, which determines the cloud emission recovered. Consequently, our sensitivity to YSOs is nonuniform across the map as the brightness of the cloud emission varies. (See Section~\ref{sec:newysos} for a discussion of the region of the brightest cloud emission, that around the early-B star \lkha 101.) For this reason, we also expect the measured fluxes of candidate YSOs to be higher than our sensitivity to an isolated point source. 
Our absolute flux sensitivity implies that we should be more sensitive to YSOs lying off of the filamentary structure; however, \Spitzer\ observations show that there are few YSOs here that are able to be detected.


\subsection{Comparison to previous YSO catalogs}
\label{sec:previouscatalogs}

The positions of extracted candidate YSOs are compared to the \Spitzer\ \citep{Gutermuthetal2009,BroekhovenFieneetal2014a} and \Herschel/PACS \citep{Harveyetal2013} YSO catalogs. We refer to YSOs by the shortest wavelength regime at which they were first identified. We detect 24 YSOs previously identified with \Spitzer\ or \Herschel/PACS (five detected by \Spitzer\ only and two identified with \Herschel/PACS only). We deem these candidate YSOs associated with a \Spitzer\ YSO or compact 70~\micron\ emission as robust protostellar objects. 

About half of the candidate YSOs identified in SCUBA-2 maps (35 out of 59) are not associated with a \Spitzer-identified or \Herschel-identified YSO. The majority of these candidate YSOs (26) are predominantly located in the bright emission near \lkha 101. Of the remaining nine candidate YSOs, two (1 and 21) are in crowded regions with multiple nearby sources, two (11 and 13) are very evident in SCUBA-2 maps and can be seen in 160~\micron\ maps, and the remaining five (43, 46, 48, 49, and 51) are elsewhere in the cloud and not very convincing visually. These five sources could very likely be merely low levels of peaks in cloud emission and not candidate YSOs. We note that it is possible that any of these candidate YSOs are actually prestellar objects and no protostar is present at their centers. Given the limitations in resolution and sensitivity above the cloud emission, interferometric observations will be required to probe the circumstellar dust emission.

Of the 22 robust YSOs that were previously detected with \Spitzer, 16 were identified as Class I,\footnote{Note that a \Spitzer-identified ``Class I'' is to be interpreted as a ``Class 0 or Class I,'' as \Spitzer\ cannot distinguish between these two spectral energy distribution classes.}, one was identified as Class F (flat spectrum), and five were identified as Class II \citep{BroekhovenFieneetal2014a}. \cite{Buckleetal2015} showed that in Taurus, the detection efficiency of the different-class YSOs declines with later classes. At 850~\micron, they recovered 88\% of the Class I YSOs, 38\% of the Class IIs, and 11\% of the Class IIIs. Auriga--Cal is 3$\times$ further away than Taurus and so our recovery of YSOs is much less, as the YSOs are even fainter (e.g., Section~\ref{sec:locations}). We recover 57\% of the Class I YSOs, 7\% of the Class Fs, 7\% of the Class IIs, and 0\% of the Class IIIs. 

We do not include the Bolocam catalog at 1.1~mm from \cite{Harveyetal2013} in our cross-matching of YSO catalogs, as these data are of lower resolution and sensitivity than our observations at the similar wavelength of 850~\micron. Bolocam has a much larger beam size (30\arcsec) compared to SCUBA-2 (14.5\arcsec\ at 850~\micron), and the mapping by Harvey et al. has noise of typically $\sim$0.07~Jy~beam$^{-1}$ (the noise is not constant in the map due to nonuniform coverage and varying observing weather conditions), which is $\sim$6$\times$\ higher than the noise in our 850~\micron\ maps. Generally, we find the Bolocam sources (14 of which fall within our SCUBA-2 coverage area) to encompass multiple YSOs identified in \Spitzer, \Herschel, and our own SCUBA-2 catalogs and/or diffuse cloud emission recovered with SCUBA-2 near those YSOs. With the higher resolution of the SCUBA-2 data, we can see the resolved substructure of the Bolocam sources, which are often somewhat offset from the YSOs (likely due to the column density of the nearby cloud emission) or at the center of multiple YSOs. The only Bolocam sources not associated with a nearby YSO are Bolocam sources 5 and 10.\footnote{The other two Bolocam sources not associated with 70~\micron\ objects from Harvey et al. are outside the SCUBA-2 coverage area.} In the SCUBA-2 map, we find that Bolocam source 5 is associated with an area of peaked emission from the cloud, although it is $\sim$30\arcsec offset from this peak. We find Bolocam source 10 to not be associated with any emission; however, it is near the region of bright emission around \lkha 101 (almost 3\arcmin\ from \lkha~101 itself), and therefore it is likely the convolution of nearby cloud emission.


\subsubsection{Bright cloud emission near \lkha 101 and implications for identifying YSOs}
\label{sec:newysos}

\begin{figure*}[h]
\includegraphics[trim=2.5cm 0cm 0.5cm 1cm, clip=True, width=7in]{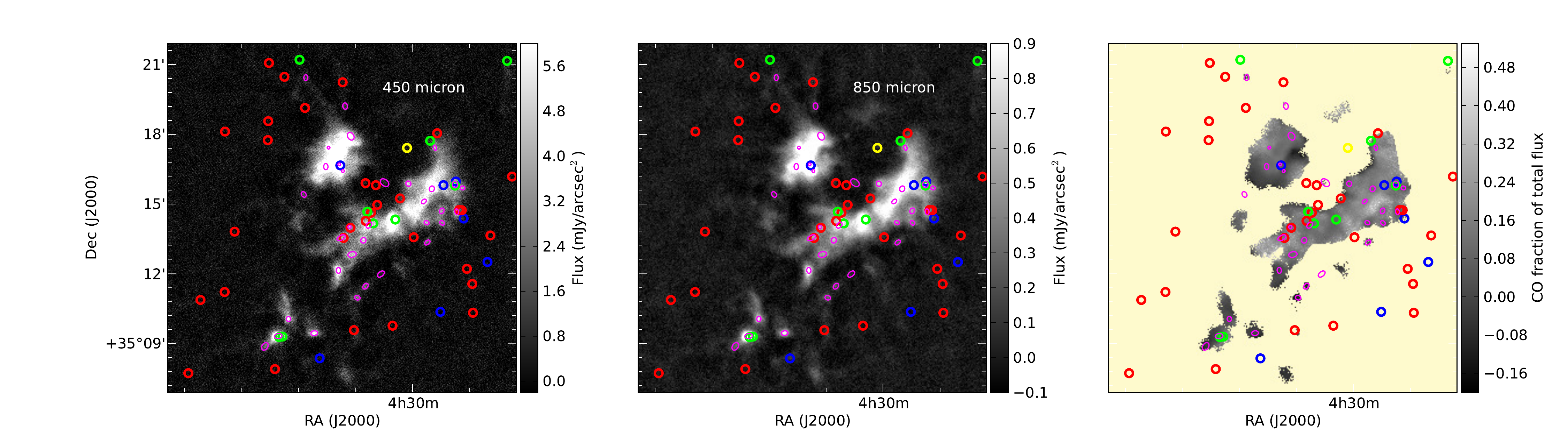}\centering
\caption{\footnotesize{The SCUBA-2 450~\micron\ (left) and 850~\micron\ (center) maps of the young cluster around \lkha 101. This is the region where the majority of candidate YSOs (i.e., those without an IR counterpart) are located. The bright cloud emission in the vicinity makes it difficult to identify YSOs at infrared wavelengths with both \Spitzer\ and \Herschel/PACS. The locations of SCUBA-2-identified candidate YSOs (magenta ellipses) are shown, with size corresponding to source FWHM. \Spitzer-identified YSOs (circles) are shown from the \citet{BroekhovenFieneetal2014a} catalog, with green, blue, red, and yellow corresponding to Class I, Class F, Class II, and Class III YSOs, respectively. Very few coincide with the location of our candidate YSOs (see discussion in text). The fractional contribution of CO to the total 850~\micron\ flux (right) is calculated from the CO-subtracted and non-CO-subtracted 850~\micron\ maps in the region around \lkha 101. We mask pixels where the 850~\micron\ flux is not detected, i.e., $< 3\sigma$ (pale yellow background) to measure the fractional contamination of CO in this region from only detected emission. Their locations coincide with lower relative CO contamination (generally levels below $\sim$20\%), since the 850~\micron\ continuum flux is higher. }\label{fig:co}} 
\vspace{-3mm}
\end{figure*}

We highlight the issue of the bright emission focusing on that near \lkha 101, particularly the arc of emission to the southwest. The bright cloud emission in this region prevents the identification of YSOs at IR wavelengths. \cite{BroekhovenFieneetal2014a} discussed the difficulty of obtaining detections of S/N $\geq$ 3 in all four IRAC bands in order to identify YSOs (their Section 3.1). A similar issue exists in the 70~\micron\ \Herschel/PACS data \citep{Harveyetal2013}, where the emission from the cloud obstructs identification of compact emission from YSOs in the vicinity of \lkha 101. At the submillimeter wavelengths presented here, the contrast between emission from the cloud and compact emission from YSOs is improved, although there is still some difficulty in isolating YSOs. We discuss here the limitations to confirming the nature of these candidate YSOs through the identification of IR emission (both in \Spitzer\ and \Herschel/PACS catalogs), constraints on their size, and the measurement of a submillimeter spectral energy distribution (SED) consistent with that of a YSO. 

In Figure~\ref{fig:co} (right), we highlight the area near \lkha 101 and compare the locations of our candidate protostellar objects to \Spitzer\  YSO catalogs \citep{Gutermuthetal2009,BroekhovenFieneetal2014a}. (There are no \Herschel/PACS--identified YSOs in this field.) As a reminder from above, the majority of candidate YSOs identified with SCUBA-2 that do not have a counterpart in the \Spitzer\ and/or \Herschel/PACS catalogs are in this region. They are mainly located in the arc of cloud emission, whereas the \Spitzer-identified YSOs are beyond the arc. There are only a handful of YSOs in the vicinity of \lkha 101 identified in both the infrared and the submillimeter catalogs. The bright cloud emission in this region, due to \lkha 101, obstructs the identification of YSOs in the infrared. In the submillimeter, however, the contrast between a YSO and the background level is higher and more favorable for YSO detection.

Typically, the presence of compact 70~\micron\ emission, associated with warmer material closer to the protostar, is used to confirm the protostellar nature of compact submillimeter sources. Compact prestellar cores, in contrast, will not be associated with compact infrared emission and will only be detected at submillimeter wavelengths. Such emission in the 70~\micron\ maps can be neither ruled out nor verified in the vicinity of \lkha 101, particularly in the nearby bright arc of cloud emission. The average flux per pixel of the arc is $\sim$0.7 Jy. This would effectively obscure compact 70~\micron\ protostellar emission, given that the median 70~\micron\ flux of robust YSOs associated with a \Herschel/PACS YSO is $\sim$1 Jy. Note that these \Herschel/PACS YSOs have been detected elsewhere in the cloud, and no YSOs are identified within 4.5\arcmin\ of \lkha 101 in the 70~\micron\ data in this region (shown in Figure~\ref{fig:co}). We therefore continue to call these sources candidate YSOs, as we cannot verify or rule out compact emission at infrared wavelengths.

The contrast between these candidate YSOs and the cloud is better at submillimeter wavelengths. Coupled with the higher resolution compared to \Herschel/SPIRE maps, this makes the peaks of compact sources easier to identify in SCUBA-2 maps. It is still difficult to disentangle cloud emission from the compact YSO emission (particularly with the lower resolution at 850~\micron). For these candidate YSOs, although we are confident in the existence of compact emission identified with \getsources, the flux associated with that compact emission remains difficult to isolate. 


\subsection{Flux measurement}
\label{sec:fluxes}

\begin{figure*}[h]
\includegraphics[trim=0cm 1.75cm 0cm 0cm, clip=True, width=6.5in]{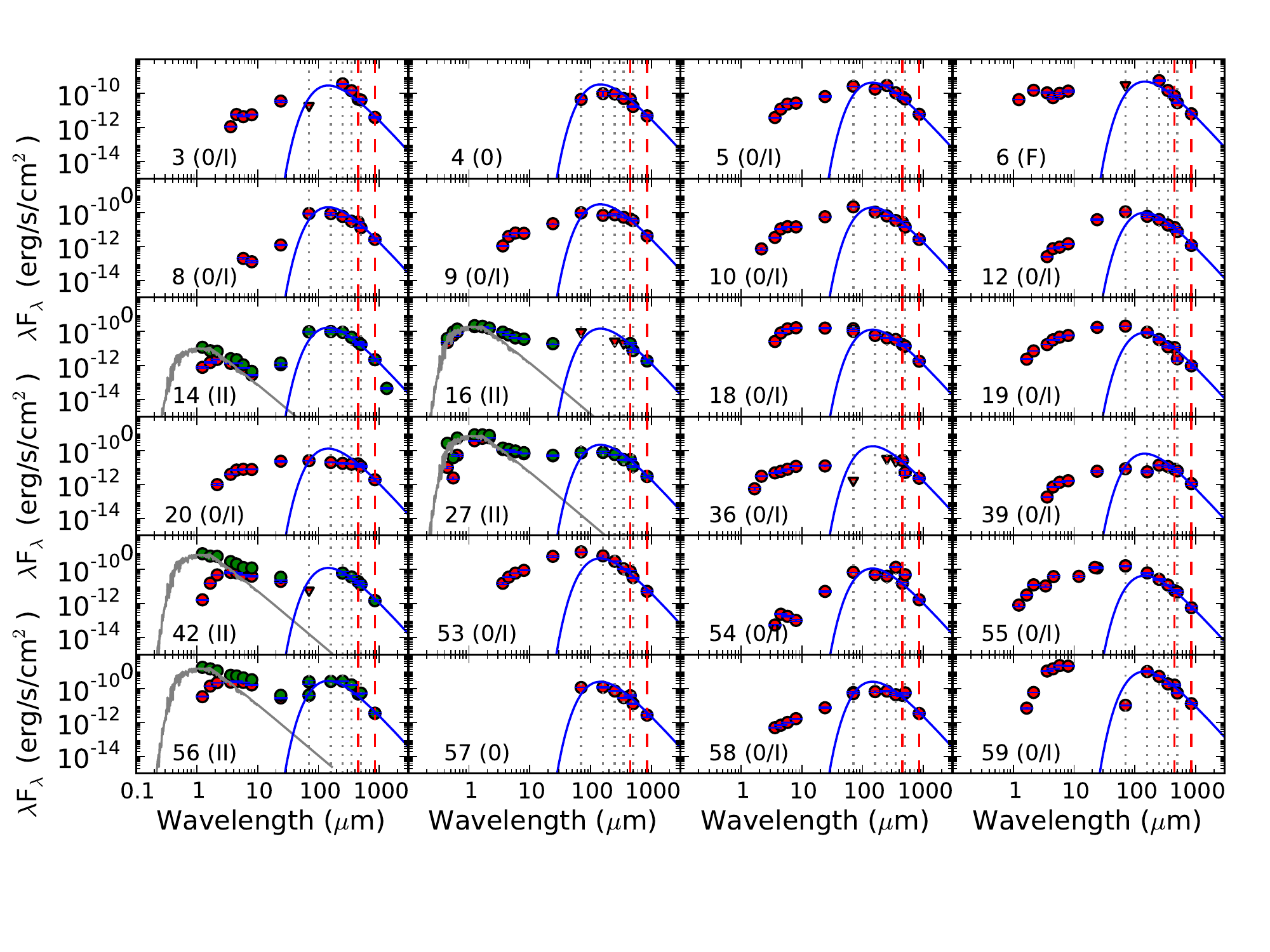}\centering
\caption{\footnotesize{SEDs for SCUBA-2 robust YSOs, i.e., those associated with a detected \Spitzer\ YSO and/or \Herschel/PACS YSO. The ID number from Table~\ref{tbl:getsources_sources} is shown in the bottom of each panel with the YSO class based on the \Spitzer\ infrared spectral slope (from \citealt{BroekhovenFieneetal2014a}) listed in parentheses. Note, however, that since \Spitzer\ cannot distinguish between Class 0 and Class I, we have labeled the \Spitzer\ Class I YSOs as Class 0/I YSOs to represent this ambiguity. Note that YSOs 4 and 57 are only identified with \Herschel/PACS in the infrared, so we use the YSO identification from \cite{Harveyetal2013} (Class 0) for these YSOs. Red circles show observed fluxes. Panels with Class II sources include a sample stellar K7 spectrum that is normalized to the infrared flux at the shortest available wavelength. The fitted stellar spectrum is used to estimate an \Av\ value and therefore the dereddened fluxes (green circles). The blue curve shows the emission from 20 K dust scaled to 450 and 850~\micron\ fluxes where available (see Section~\ref{sec:masses}) and listed in Table~\ref{tbl:sedanal}. The red dashed lines mark the SCUBA-2 wavelengths at 450 and 850~\micron, and the black dotted lines mark the \Herschel/SPIRE wavelengths at 250, 350, and 500~\micron.\label{fig:seds}}}
\end{figure*}

\begin{figure*}[h]
\includegraphics[trim=0cm 0.93cm 0cm 0cm, clip=True, width=6.5in]{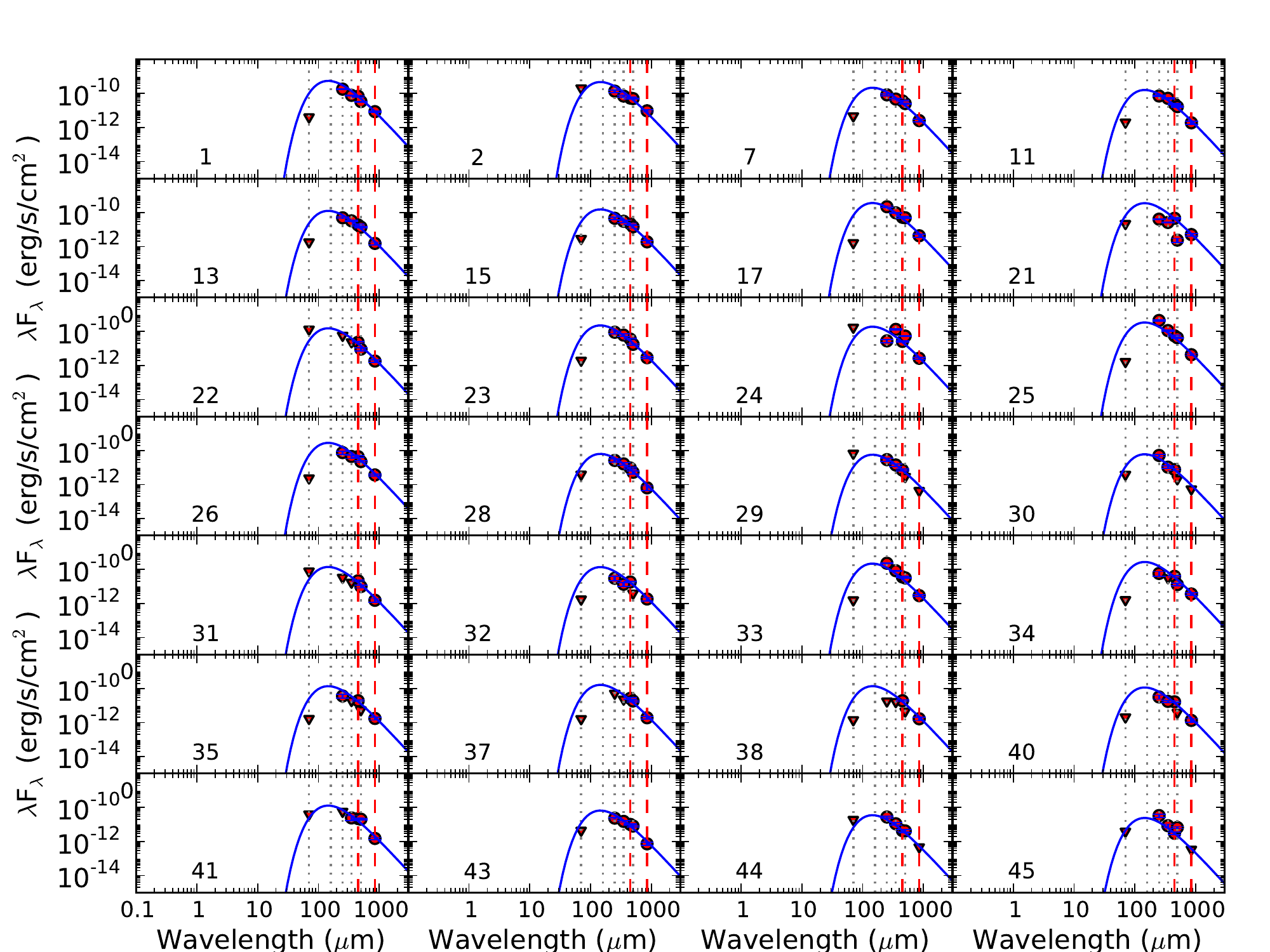}\centering \\
\vspace{-0.05cm}
\includegraphics[trim=0cm 9.4cm 0cm 0.95cm, clip=True, width=6.5in]{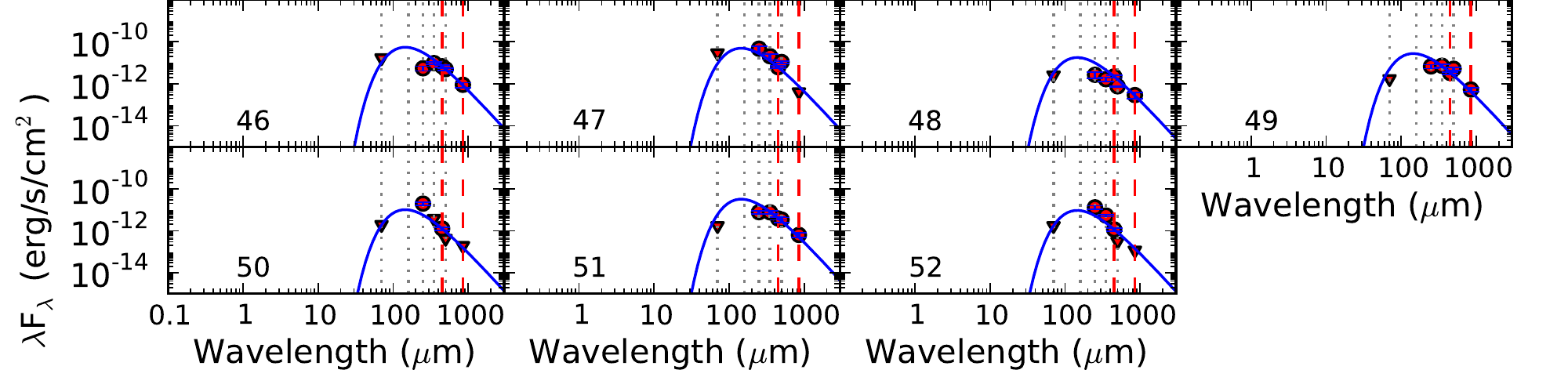}\centering
\caption{\footnotesize{SEDs for SCUBA-2 candidate YSOs not associated with \Spitzer-identified or \Herschel/PACS-identified YSOs (probably due to observational constraints in the region rather than the absence of compact IR flux; see Section~\ref{sec:newysos}). The ID from Table~\ref{tbl:getsources_sources} is shown in the bottom of each panel. The blue curve shows the emission from 20 K dust scaled to 450 and 850~\micron\ fluxes where available (see Section~\ref{sec:masses}) and listed in Table~\ref{tbl:sedanal}. The red dashed lines mark the SCUBA-2 wavelengths at 450 and 850~\micron, and the black dotted lines mark the \Herschel/SPIRE wavelengths at 250, 350, and 500~\micron.\label{fig:seds2}}}
\end{figure*}

We include the fluxes of YSOs measured in maps from multiple infrared and submillimeter instruments, all with different resolutions affecting their appearance. In addition to the effect of resolution, the intrinsic spatial scale sampled of the emission from a YSO depends on the wavelength at which it is observed. At infrared wavelengths, the emission is expected to be more compact than at submillimeter wavelengths, as it originates from the more central material that is warmed by the protostar. The infrared emission can be fit as a point source, even in the 6\arcsec\ resolution \Herschel/PACS 70~\micron\ images. YSO emission at submillimeter wavelengths, however, despite the lower resolution (7.5\arcsec\ and 14.5\arcsec\ for SCUBA-2), must be fit with allowance for a more extended profile, to account for cooler dust emission from the envelope that can have a size up to  10,000 au ($\sim$22\arcsec\ at Auriga--Cal's distance). For each wavelength, therefore, the method we use to determine the flux of the emission associated with the YSO depends on the resolution of the maps and the contrast between the cloud and YSO emission. This is preferable to extracting the flux in the same manner at all wavelengths, as such an analysis would require essentially considering all emission at the resolution corresponding to the lowest-resolution map (in this case, the 36\arcsec\ resolution of the 500~\micron\ \Herschel/SPIRE maps). Such fluxes would be essentially consistent pixel to pixel; however, we would lose the benefits of the higher-resolution maps in better isolating compact sources and would therefore result in more cloud emission contributing to all fluxes at all wavelengths. We therefore measure the fluxes at each of the PACS, SPIRE, and SCUBA-2 wavelengths using the techniques that have been found to be optimal for the corresponding instrument. For each instrument, we describe how the fluxes are measured and how that technique has accounted for the cloud emission in order to isolate it from the YSOs.

The \Herschel/PACS fluxes for YSO candidates associated with a \Herschel/PACS--identified YSOs are adopted from \citet{Harveyetal2013} (measured using the \textit{c2dphot} package developed for the Spitzer Legacy c2d program: \citealt{Harvey2006,Evans2007}). These data are very straightforward for identifying compact sources: the 70~\micron\ map is essentially composed of point-like sources with some cloud emission only in the region around \lkha 101 (albeit very bright). The 160~\micron\ map has large-scale cloud emission, but it is faint relative to the bright YSOs, which still appear point-like. We also include some 70~\micron\ flux upper limits for SCUBA-2 YSO candidates from this work that are not associated with an infrared YSO (either \Herschel/PACS or \Spitzer). This is to help illustrate how any compact 70~\micron\ emission is indistinguishable from bright cloud emission in the vicinity of the early-B star \lkha 101 and represent such limitation on the spectral energy distributions (SEDs). For these candidate YSOs, we measure the upper limit on the possible 70~\micron\ flux as the total flux within a 20\arcsec\ aperture centered on each candidate YSO's position.

The \Herschel/SPIRE fluxes are extracted from the source properties measured by \getsources\ (see Section~\ref{sec:getsources}). This characterization of the flux is straightforward as \getsources\ has been designed for and tested on \Herschel\ data. The compact emission is disentangled from the cloud emission by the spatial decompositions and background modeling done within \getsources.

For SCUBA-2 fluxes, aperture photometry is used rather than the properties extracted by \getsources. Although \getsources\ is a rigorous algorithm to disentangle emission from nearby sources and the cloud to identify sources (e.g., Figure~\ref{fig:co}), the SCUBA-2 beam shape is very different from the \Herschel\ beam, particularly at 450~\micron\  where about half the beam power is concentrated in the secondary beam that extends beyond the primary 7.5\arcsec\ component \citep{Dempseyetal2013}. The primary beam of both the 450~\micron\ and 850~\micron\ beams is still relatively Gaussian \citep{Dempseyetal2013} allowing \getsources\ to identify Gaussian-like sources with spatial decompositions. More subtle differences from the \Herschel\ beam shapes in the larger-scale components of the JCMT beam shapes, however, complicate using the flux measurements by \getsources\ for SCUBA-2 data. Using aperture photometry on SCUBA-2 maps, however, is well-tested and better understood and at least provides a more-straightforward comparisons to measurements made on similar datasets \citep{Dempseyetal2013,Buckleetal2015,Pattleetal2015}. We therefore also use aperture photometry to measure 450~\micron\ and 850~\micron\ fluxes. Fluxes are measured using apertures with sizes 2$\times$FWHM along major and minor axes and aligned to the position angle measured with \getsources, similar to that done by \cite{Pattleetal2015}. The fluxes are included in Table~\ref{tbl:getsources_sources}. Note that many of the YSOs are colocated with bright cloud emission (particularly in the region around \lkha 101) and/or have nearby YSOs. (These are marked in Table~\ref{tbl:getsources_sources}.) The aperture fluxes included here are therefore likely an over estimate of the true flux. 

We collect available photometry from 2MASS, \Spitzer, and \Herschel/PACS to complement the SCUBA-2 and \Herschel/SPIRE photometry measured here. The SEDs for SCUBA-2 candidate YSOs associated with  \Spitzer-identified YSOs are shown in Figure~\ref{fig:seds} along with their \Spitzer\ class identification. The near-infrared photometry of Class II YSOs is more likely to trace the photosphere of the central star than the enshrouded Class I and earlier YSOs, and so we have included a sample K7 stellar spectrum normalized to the near-infrared flux at the shortest available wavelength for these YSOs. SEDs for candidate YSOs without an infrared counterpart are shown in Figure~\ref{fig:seds2}.


\subsection{CO contamination at the location of compact sources}
\label{sec:source_CO}

Figure~\ref{fig:co} shows a map of the fractional contribution of CO to the total flux in the region around \lkha 101 where both dust and CO are detected. Overall, we find the locations of compact sources to coincide with areas of lower relative CO contamination. Specifically, the measured CO contamination in NGC 1529 is below 20\% at the locations of compact sources (see Section~\ref{sec:getsources} and Figure~\ref{fig:co}, right) and less than 50\% elsewhere. As the total 850~\micron\ emission peaks at source locations, the relative contribution of CO is lower for these objects. Additionally, the CO line saturates in bright areas, which consequently limits its relative contribution to the 850~\micron\ map even more. Ultimately, we do not find that outflows from nearby YSOs are significantly contaminating the 850~\micron\ fluxes of other YSOs, and we expect that the contamination of 850~\micron\ fluxes of YSOs elsewhere in the cloud is also $<$20\%.


\subsection{Masses}
\label{sec:masses}

The circumstellar masses, $M$, of submillimeter-detected robust and candidate YSOs are calculated from the 450 and 850~\micron\ fluxes, where available, using  
\begin{equation}\label{eq:dustMass}
M_{} = \frac{F_{dust} D^2}{\kappa_{\nu}B_{\nu}(T_{\rm dust})}
\end{equation}
where $D$ is the distance to the source, $\kappa_{\nu}$ is the opacity of the dust grains, and $B_{\nu}(T_{\rm dust})$ is the Planck function for temperature, $T_{\rm dust}$. The opacity is assumed to be \newline $\kappa_{\nu}=0.1(\nu/$1000~GHz$)^\beta$ cm$^2$ g$^{-1}$ \citep{Beckwithetal1990}. Note that the opacity relation includes an assumed dust-to-gas ratio of 1:100, and therefore $M$ represents the total dust+gas mass of the circumstellar material (disk and
envelope). Therefore, the submillimeter mass will be dominated by the envelope for younger (i.e., Class 0/I) sources, whereas the submillimeter mass for more-evolved sources (i.e., Class IIs) will reflect material that remains in the circumstellar disk after the envelope is dissipated. Our observations are sensitive to the YSOs with the most massive circumstellar material, i.e., the Class 0/Is that are still surrounded by an envelope, as demonstrated by the majority of detected YSOs identified with \Spitzer\ YSOs being Class I/F objects (17 out of 22), with much fewer associated with Class II objects (5 out of 22; Section~\ref{sec:previouscatalogs}). 

We assume the fiducial parameters of $T=20$ K and $\beta$ = 1 when calculating circumstellar masses from the submillimeter thermal dust emission of all candidate YSOs \citep{AndrewsWilliams2005, AndrewsWilliams2007, MannWilliams2009a, Mohantyetal2013, Williamsetal2013, Mannetal2014, Mannetal2015}. The expected uncertainty when assuming this single temperature and $\beta$ for all YSOs is a factor of a few \citep{AndrewsWilliams2005}. This provides a standard for comparison with other YSO populations and works in the literature and is also reasonable for YSOs of different classes. We expect objects of earlier class to be slightly cooler ($\sim$15 K; \citealt{Youngetal2003}), shielded by their envelope, and objects of later class to be slightly warmer. We note, however, that we expect a higher contribution from the envelope material of the Auriga--Cal YSOs than from the YSOs studied in the works listed above. Auriga--Cal's is more distant compared to Taurus and $\rho$ Oph, resulting in more of the outer envelope being included in the beam, and the measured masses in Orion A are from interferometric observations that filter out emission on larger spatial scales.

To test whether these values are representative of our (candidate and robust) YSOs, we perform a least-squares fit of Equation~\ref{eq:dustMass} for the 19 YSOs that have $\geq$ 3$\sigma$ detections at all six wavelengths of 160~\micron\ (PACS), 250~\micron, 350~\micron, 500~\micron\ (SPIRE), 450~\micron, and 850~\micron\ (SCUBA-2). This sample ensures that we are using the most robust YSOs (i.e., YSOs identified in SCUBA-2 maps with an IR counterpart and strong detections with both \Herschel\ and SCUBA-2) to test the values, which we then apply to all candidate and robust YSOs. We use the least-squares minimization package MPFIT \citep{Markwardt2009} to fit temperature, $T$, and $\beta$ to the SED of each YSO at wavelengths between 160 and 850~\micron\ (inclusive). More accurate fitting of temperatures for our YSOs would require more detailed SED modeling to account for YSO geometry and the dust emission from different regions of the circumstellar material (and their respective temperatures), and such an analysis is beyond the scope of this work. The wavelength range included here models across the peak of the SED of the thermal emission from the coldest dust. The wavelength cutoff of 160~\micron\ is used, as emission at wavelengths $<$ 100~\micron\ is optically thick at all radii in disks \citep{Beckwithetal1990}, {but note that emission  at 160~\micron\ is expected to be optically thick at most regions of the disk.}  This boundary, however, is also consistent with the cutoffs used to measure temperature and $\beta$ for clumps and young YSOs \citep[e.g.,][]{Youngetal2003,Sadavoy2013,Sadavoyetal2014}, which are less dense than disks and optically thin at 160~\micron. It is therefore useful to include this wavelength for envelope-dominated YSOs. The fractional contribution of optically thick emission to the total flux decreases with increasing wavelength; for this reason, the submillimeter emission provides the most important anchor for these SED fits. The fitted temperatures and $\beta$ values for these YSOs are listed in Table~\ref{tbl:chisqfits}. This table also includes the reduced $\chi^2$ value, $\chi^{2}_{\rm{reduced}}$, which is the $\chi^2$ value divided by the number of degrees of freedom. The average fitted temperature for these 19 YSOs is $17 \pm 6$ K and the average fitted $\beta$ is 1.1 $\pm$ 0.6 (with standard deviations quoted as uncertainties). These values are consistent with the fiducial temperature (20 K) and $\beta$ (1) assumed (and the lower temperatures  found for earlier-class objects), and therefore we proceed to measure the masses for our entire sample assuming these values.

\begin{deluxetable}{cccc}
\tablecolumns{4}
\tablewidth{3.3in}
\tablecaption{SED $\chi^2$ minimization fits } 
\tablehead{
\ch{\textcolor{white}{ID} ID \textcolor{white}{ID}}  & \ch{Temperature (K)} & \ch{$\beta$} & \ch{$\chi^{2}_{\rm{reduced}}$} }
\startdata
       4 &                   17 &          0.8 &     3.9                  \\ 
       5 &                   12 &          2.0 &     5.2                  \\ 
       8 &                   19 &          0.6 &     1.8                  \\ 
       9 &                   13 &          1.1 &     3.6                  \\ 
      10 &                   21 &          0.5 &     1.4                  \\ 
      12 &                   19 &          0.8 &     0.7                  \\ 
      14 &                   13 &          1.8 &     0.1                  \\ 
      18 &                   16 &          0.8 &     3.9                  \\ 
      19 &                   30 &          0.9 &     3.3                  \\ 
      20 &                   17 &          0.1 &     4.6                  \\ 
      27 &                   20 &          0.5 &     2.6                  \\ 
      39 &                    9 &          1.7 &     0.04                  \\ 
      53 &                   21 &          1.4 &     2.2                  \\ 
      54 &                   11 &          2.0 &     58                  \\ 
      55 &                   24 &          0.9 &     0.4                  \\ 
      56 &                   13 &          2.0 &     9.5                  \\ 
      57 &                   25 &          0.5 &     4.5                  \\ 
      58 &                   10 &          2.0 &     21                  \\ 
      59 &                   21 &          1.3 &     5.3                  \\ 
\enddata
\label{tbl:chisqfits}
\tablecomments{\small Results of the $\chi^2$ minimization fits to the SEDs at long wavelengths ($\geq$ 160~\micron) for robust YSOs with $\geq3 \sigma$ flux detections at all seven wavelengths of \Herschel/PACS, \Herschel/SPIRE, and SCUBA-2 (see description in Section~\ref{sec:masses}). The values for temperature and $\beta$ of the best fits are listed, along with the reduced $\chi^2$ value for each fit ($\chi^2$ divided by the number of degrees of freedom). The average ($\pm$ the standard deviation) of the fitted temperatures and $\beta$ values are 17 $\pm$ 6 K and 1.1 $\pm$ 0.6, respectively. These are consistent with the fiducial values typically used to measure circumstellar dust masses and what we adopt for our own circumstellar mass measurements.  }
\end{deluxetable}

The fits are generally good, with the exception of the 450~\micron\ flux, which is often higher in comparison to both the minimization fits (and is the cause for the higher $\chi^{2}_{\rm{reduced}}$ values aside from YSO 54, where the 350 and 500~\micron\ fluxes are largely discrepant from the rest of the observed SED) and with an interpolation between 350 and 500~\micron\ in general. We expect, as noted in Section~\ref{sec:fluxes}, that the fluxes for many of our YSOs will be contaminated by cloud emission and/or nearby sources. Such contamination inflating the fluxes could result in the 450~\micron\ flux discrepancy and also decrease the measured $\beta$ value with an inflated 850~\micron\ flux. Indeed, the spectral slope between the 450 and 850~\micron\ is steeper than what is expected from 20 K dust with $\beta$=1 and suggests a larger $\beta$ value or hotter temperature. A larger $\beta$ value would be consistent with contamination from cloud emission, typically having a $\beta$ value of $\sim$1.7, as well as with $\beta$ values measured toward early-class protostars (e.g., \citealt{Chenetal2016}).

\begin{deluxetable}{lcccc}
\tablecolumns{4} 
\tablewidth{0pc} 
\tablecaption{Circumstellar masses}
\tablehead{
\ch{ID}  & \ch{IR Classification} & \ch{$M_{}$}\\
\ch{}    & \ch{}                  & \ch{(\Msolar)}\\
\ch{(1)} & \ch{(2)}               & \ch{(3)}}
\startdata
       1 &        \nodata &   1.37 $\pm$ 0.10 \\       
       2 &        \nodata &   1.17 $\pm$ 0.11 \\       
       3 &      Class 0/I &   0.73 $\pm$ 0.06 \\       
       4 & Class 0 (PACS) &   0.85 $\pm$ 0.07 \\       
       5 &      Class 0/I &   1.08 $\pm$ 0.08 \\       
       6 &        Class F &   1.23 $\pm$ 0.12 \\       
       7 &        \nodata &   0.54 $\pm$ 0.05 \\       
       8 &      Class 0/I &   0.51 $\pm$ 0.05 \\       
       9 &      Class 0/I &   0.75 $\pm$ 0.07 \\       
      10 &      Class 0/I &   0.49 $\pm$ 0.05 \\       
      11 &        \nodata &   0.40 $\pm$ 0.04 \\       
      12 &      Class 0/I &   0.24 $\pm$ 0.03 \\       
      13 &        \nodata &   0.31 $\pm$ 0.03 \\       
      14 &       Class II &   0.42 $\pm$ 0.04 \\       
      15 &        \nodata &   0.38 $\pm$ 0.04 \\       
      16 &       Class II &   0.36 $\pm$ 0.04 \\       
      17 &        \nodata &   0.90 $\pm$ 0.09 \\       
      18 &      Class 0/I &   0.32 $\pm$ 0.03 \\       
      19 &      Class 0/I &   0.21 $\pm$ 0.03 \\       
      20 &      Class 0/I &   0.32 $\pm$ 0.03 \\       
      21 &        \nodata &   0.87 $\pm$ 0.08 \\       
      22 &        \nodata &   0.38 $\pm$ 0.04 \\       
      23 &        \nodata &   0.56 $\pm$ 0.05 \\       
      24 &        \nodata &   0.47 $\pm$ 0.04 \\       
      25 &        \nodata &   0.84 $\pm$ 0.07 \\       
      26 &        \nodata &   0.69 $\pm$ 0.06 \\       
      27 &       Class II &   0.54 $\pm$ 0.05 \\       
      28 &        \nodata &   0.16 $\pm$ 0.02 \\       
      29 &        \nodata &   0.14 $\pm$ 0.02 \\       
      30 &        \nodata &   0.15 $\pm$ 0.02 \\       
      31 &        \nodata &   0.35 $\pm$ 0.04 \\       
      32 &        \nodata &   0.34 $\pm$ 0.04 \\       
      33 &        \nodata &   0.55 $\pm$ 0.04 \\       
      34 &        \nodata &   0.69 $\pm$ 0.06 \\       
      35 &        \nodata &   0.34 $\pm$ 0.03 \\       
      36 &      Class 0/I &   0.44 $\pm$ 0.04 \\       
      37 &        \nodata &   0.40 $\pm$ 0.04 \\       
      38 &        \nodata &   0.34 $\pm$ 0.04 \\       
      39 &      Class 0/I &   0.16 $\pm$ 0.02 \\       
      40 &        \nodata &   0.28 $\pm$ 0.03 \\       
      41 &        \nodata &   0.32 $\pm$ 0.03 \\       
      42 &       Class II &   0.30 $\pm$ 0.03 \\       
      43 &        \nodata &   0.16 $\pm$ 0.02 \\       
      44 &        \nodata &   0.09 $\pm$ 0.01 \\       
      45 &        \nodata &   0.06 $\pm$ 0.01 \\       
      46 &        \nodata &   0.14 $\pm$ 0.02 \\       
      47 &        \nodata &   0.12 $\pm$ 0.02 \\       
      48 &        \nodata &   0.04 $\pm$ 0.01 \\       
      49 &        \nodata &   0.07 $\pm$ 0.01 \\       
      50 &        \nodata &   0.03 $\pm$ 0.00 \\       
      51 &        \nodata &   0.08 $\pm$ 0.01 \\       
      52 &        \nodata &   0.02 $\pm$ 0.00 \\       
      53 &      Class 0/I &   1.12 $\pm$ 0.10 \\       
      54 &      Class 0/I &   0.28 $\pm$ 0.02 \\       
      55 &      Class 0/I &   0.11 $\pm$ 0.01 \\       
      56 &       Class II &   0.69 $\pm$ 0.06 \\       
      57 & Class 0 (PACS) &   0.62 $\pm$ 0.07 \\       
      58 &      Class 0/I &   0.62 $\pm$ 0.05 \\       
      59 &      Class 0/I &   2.48 $\pm$ 0.21 \\       
\enddata
\label{tbl:sedanal}
\tablecomments{\small Column (1): ID from Table~\ref{tbl:getsources_sources}. Column (2): classification based on the \Spitzer\ infrared spectral slope (from \citealt{BroekhovenFieneetal2014a}) for robust YSOs associated with a detected \Spitzer\ YSO and/or \Herschel/PACS YSO. YSOs only detected with PACS and not \Spitzer\ are noted with `PACS' in parentheses and use the YSO identification from \cite{Harveyetal2013}. Note that \Spitzer\ cannot distinguish between Class 0 and Class I, so we have used Class 0/I instead of Class I, as noted in c2d and related works. Column (3): circumstellar mass measured using Equation~\ref{eq:dustMass} and assuming $T_{dust}$ = 20 K and $\beta =$ 1 (see Section~\ref{sec:masses}).}
\end{deluxetable}

\begin{figure*}[h]
\includegraphics[trim=0cm 0cm 0cm 1.34cm, clip=True, height=2.5in]{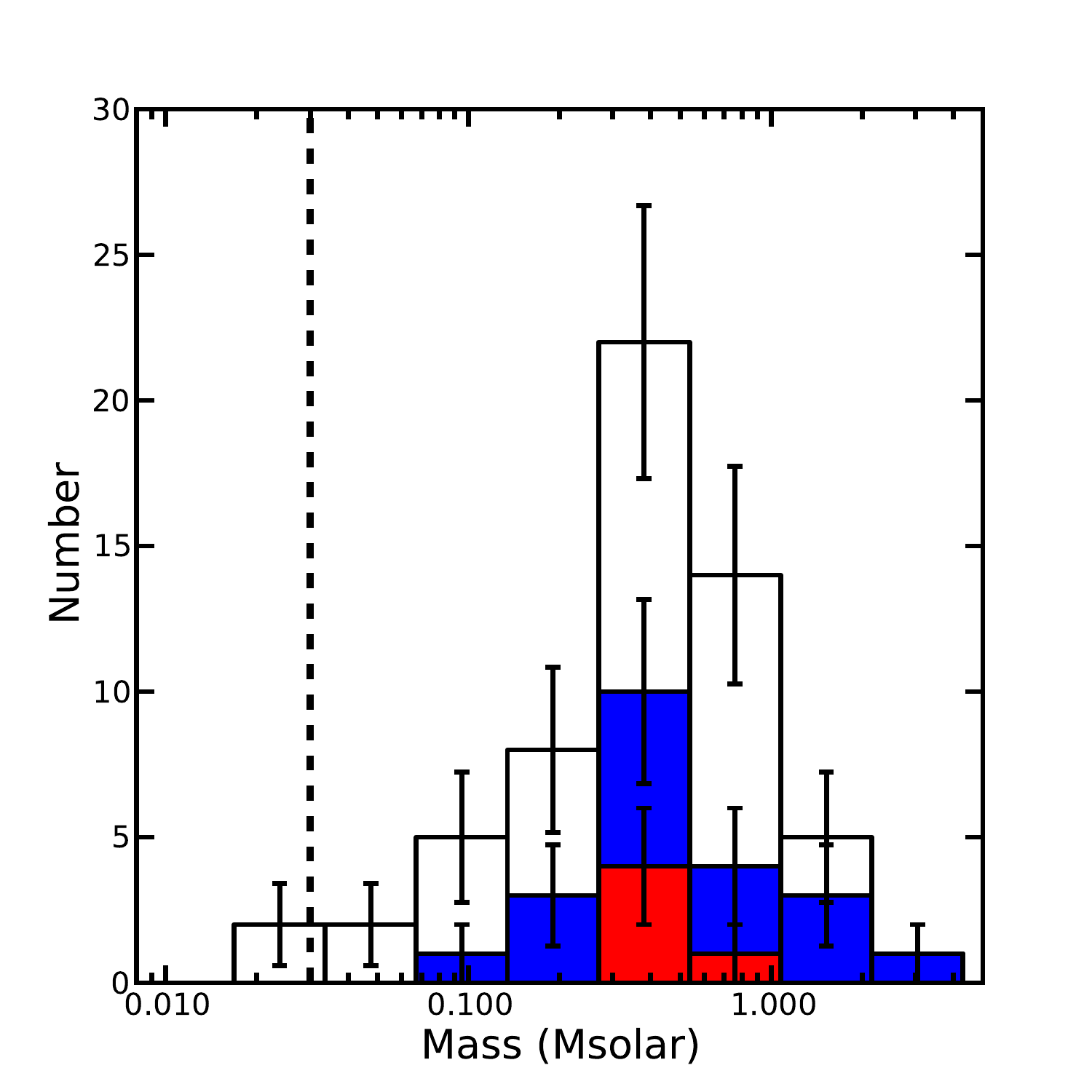}\centering
\caption{\footnotesize{Circumstellar (disk + envelope) mass distribution of calculated masses. The distribution for all YSOs, both candidate and robust, is shown in white. The distribution shown in blue includes only robust YSOs with a \Spitzer\ counterpart, and the Class II YSOs are shown in red. The JCMT GBS observations are sensitive to the high-mass end of the mass distribution. The dashed vertical line marks our 3$\sigma$ limit at 850~\micron absent of bright cloud emission at 0.03~\Msolar\ assuming $T =$ 20 K and $\beta =$ 1. (Mass measurements leftward of this line are due to 450~\micron\ fluxes.)}\label{fig:comparediskmasses}}
\end{figure*}

We calculate the circumstellar masses from the SCUBA-2 submillimeter wavelengths where the dust emission is most optically thin (i.e., the contribution from optically thick emission to the total flux is lower at longer wavelengths). 
For candidate YSOs detected at both wavelengths, we report the weighted mean of the mass calculated from each flux at 450 and 850~\micron. Mass uncertainties are derived from the flux uncertainties. All masses are listed in Table~\ref{tbl:sedanal}, and the distribution of these measurements is shown in Figure~\ref{fig:comparediskmasses}. The similar shape of the distribution for candidate + robust YSOs (white) and robust YSOs with a \Spitzer\ counterpart (blue) supports that each distribution samples similar objects, that is, that our candidate YSOs are indeed protostellar. The Class II distribution is also similar to the distribution for the whole sample, suggesting that they are drawn from similar populations. Indeed, all Spitzer Class II YSOs detected in this work are suggested to be earlier-stage objects upon recalculation of the IR spectral slope including the fluxes measured at the longer PACS wavelengths (Harvey et al. 2013). Therefore, it is likely that these are earlier-stage objects still having some envelope that are identified as Class II SEDs with \Spitzer\ observations due to viewing geometry (more pole-on).

Figure~\ref{fig:comparediskmasses} also highlights that the JCMT GBS observations are sensitive to only the high-mass end of the circumstellar mass distribution. In the absence of bright cloud emission, our 3$\sigma$ limit at 850~\micron (assuming the temperature and $\beta$ above) corresponds to 0.03~\Msolar, which is three times the minimum mass solar nebula (MMSN). This is comparable to the median disk mass of 0.025~\Msolar\ for Class Is in Taurus \citep{AndrewsWilliams2005}, and so it is unsurprising that we recover about half (57\%) of the Class Is in Auriga--Cal (Section~\ref{sec:previouscatalogs}). In contrast, only 10\% and 17\% of YSOs (all classes) have measured circumstellar masses $>$0.03~\Msolar\ in $\rho$ Ophiuchi \citep{AndrewsWilliams2007} and Taurus-Auriga, respectively. This fraction is similar to the 13\% of \Spitzer-identified YSOs in Auriga--Cal that we recover with SCUBA-2. 
There are, however, more masses that are $> 0.1~$\Msolar\ in Auriga--Cal than are measured in Taurus and $\rho$ Ophiuchi. These clouds, however, are much closer (140 and 150 pc, respectively) than Auriga--Cal; consequently, the masses measured in this work may be artificially higher due to more mass from surrounding material being included in a single beam. Furthermore, Auriga--Cal has a higher fraction of Class I objects than these clouds \citep{BroekhovenFieneetal2014a}, so it should have more objects at a younger stage and therefore with more circumstellar material. The Orion star-forming complex is at a similar distance as Auriga--Cal; however, the regions studied within it are either much older, and therefore have more-evolved YSOs with less circumstellar material (e.g., $\sigma$ Orionis; \citealt{Williamsetal2013}), or were observed with an interferometer, which filters out more of the larger-scale envelope emission (e.g., the Trapezium cluster: \citealt{MannWilliams2010,Mannetal2014}; and the NGC 2024 cluster: \citealt{Mannetal2015}). It would be interesting to determine whether disk mass distribution near the early-B star \lkha 101 was truncated, as is observed for disks in proximity to the O-star $\theta^1$Ori C in Orion A North \citep{MannWilliams2009a,Mannetal2014}. It is difficult, however, to investigate such a measurement in this region with single-dish observations given the bright background cloud emission coupled with the small angular proximity of YSOs. 

We show the SEDs for robust YSOs (those with counterparts in the \Spitzer\ YSO catalog \citep{BroekhovenFieneetal2014a} and/or the \Herschel/PACS YSO catalog) in Figure~\ref{fig:seds} and for the candidate YSOs in Figure~\ref{fig:seds2}. The emission profile of the calculated masses (by rearranging Equation~\ref{eq:dustMass}) is shown against the observed fluxes for all objects. It can be seen in Figures~\ref{fig:seds} and \ref{fig:seds2} that the shape of this blue curve generally follows the observed submillimeter fluxes well. This agreement confirms that our parameter assumptions are a reasonable representation for our sample of YSOs. For some robust YSOs (e.g., 3, 42, and 36) with far-IR detections, Figure~\ref{fig:seds} shows that the blue curve traces higher fluxes than what is observed at far-IR wavelengths. This discrepancy highlights that there is a higher fraction of optically thick dust at these wavelengths, as well as multiple dust temperatures contributing to said flux. The upper limits on 70~\micron\ fluxes of candidate YSOs in Figure~\ref{fig:seds2} are consistent with the blue curve, similar to the SEDs for robust YSOs, and exhibit cases where the blue curve exceeds the 70~\micron\ flux upper limits. As for robust YSOs with similar behavior, this is likely due to optically thick emission at these wavelengths. Furthermore, recall that the YSOs with 70~\micron\ flux upper limits are those in bright regions of the cloud (Section~\ref{sec:newysos}) where 450 and 850~\micron\ fluxes are also susceptible to more contamination from nearby cloud emission and YSOs (Section~\ref{sec:fluxes} and Table~\ref{tbl:getsources_sources}), and so the masses for theses sources (and the corresponding blue lines in Figures~\ref{fig:seds} and \ref{fig:seds2}) are more likely to be overestimated.


\subsection{Comparison with Orion A}
\label{sec:Orion}

Infrared observations with \Spitzer\ suggest that Auriga--Cal is forming 15--20 times fewer stars than Orion A \citep{BroekhovenFieneetal2014a}. Auriga--Cal and Orion A are the most distant GBS clouds, both at $\sim$450 pc, with similar physical resolution and emission sensitivity. Furthermore, they share similar filamentary morphology (in contrast to Orion B, for example, which more resembles pockets of high-density material), thus making Orion A an ideal and intriguing region for comparison. The SCUBA-2 coverage of the clouds is different than that of the \Spitzer\ surveys. We consider only the fractional difference of \Spitzer-identified YSOs within the SCUBA-2 coverage, of which there are 1309 in Orion A and 123 in Auriga--Cal \citep{BroekhovenFieneetal2014a}. (The count for \Spitzer-identified YSOs in Orion A is limited to those from \citealt{Megeathetal2012} with detections in photometry of all four IRAC bands, as this is more similar to the source list of \citealt{BroekhovenFieneetal2014a}; see discussion in their Section~2.3.2.) Therefore, there are 11 times more \Spitzer-identified YSOs in Orion A than in Auriga--Cal within the areas observed by SCUBA-2.

We investigate how this fraction extends to YSOs observable in the submillimeter by comparing the number of compact sources in Orion A and Auriga--Cal, as observed with SCUBA-2. We use a \getsources\ extraction from JCMT GBS observations of Orion A \citep{Laneetal2016} to ensure that the comparison is unbiased by source identification algorithms. We take the final catalog produced by \getsources\ and perform the same cuts on flux and geometry as we did for sources in Auriga--Cal (see Section~\ref{sec:getsources}), including the cuts on the ratio between the sizes and aspect ratios measured individually at 450 and 850~\micron, to identify a total of 539 compact sources (or 300 that satisfy the criteria at both 450 and 850~\micron). We do not do a visual vetting and source catalog comparison for the Orion A compact source list for this purpose, and so we compare this total number of compact sources in Orion A (539 at either wavelength or 300 at both wavelengths) with the number of compact sources identified in Auriga--Cal (79 at either wavelength or 44 at both wavelengths), rather than the number of candidate YSOs (59). There are $\sim$7 times more submm compact sources in Orion A (for both the compact sources that satisfy the criteria at either wavelength and the compact sources that satisfy the criteria at both wavelengths).

The ratio of compact submillimeter sources observed with SCUBA-2 in Auriga--Cal relative to Orion A (7 times fewer) is similar to the ratio of \Spitzer-identified YSOs within the SCUBA-2 coverage area of the two clouds (11 times fewer in Auriga--Cal). The consistency of these ratios shows that the disparity between the two clouds of the number of star-forming objects observed in the infrared extends to the submillimeter. Recall that this difference in star formation is attributed to the difference in mass at high density between the two clouds \citep{Ladaetal2009} of about an order of magnitude. The similar ratios for the embedded and the nonembedded populations in Auriga--Cal and Orion A suggest that there has been no significant difference in the relative star formation rates over the Class II lifetime, as there is not a larger fraction of very young YSOs. This implies that Auriga--Cal is not expected to be as productive as Orion A for the foreseeable future, if ever.


\section{Summary}
\label{sec:summary}

We analyzed the SCUBA-2 observations of Auriga--Cal at 450 and 850~\micron\ as part of the JCMT GBS. We identify 79 compact sources in the SCUBA-2 maps using the \getsources\ algorithm and find 59 objects that we identify as candidate YSOs on the basis of the SCUBA-2 data alone. The majority of these candidate YSOs are colocated with cloud emission, consistent with observations of YSOs along natal filaments \citep{Andreetal2010}. More candidate YSOs are identified at 450~\micron\ than at 850~\micron, in part due to the higher resolution. The YSOs, however, are also brighter at 450~\micron, and the contrast there is increased between the compact emission and the cold background emission, as these candidate YSOs are generally in the brightest areas of cloud emission, predominately around \lkha 101. Given the complexity and richness of the \lkha 101 cluster, the only way to get a census of YSO circumstellar masses in this region is with spatially filtered interferometric observations.

We compared our catalog of candidate YSOs in SCUBA-2 maps with catalogs of \Spitzer\ and \Herschel/PACS YSOs. Approximately half of the SCUBA-2 candidate YSOs (24 out of 59) are associated with an infrared-identified YSO and therefore are deemed to be robust protostellar objects. The majority of the remaining SCUBA-2 candidate YSOs are in areas of bright background emission, mainly the arc of emission near \lkha 101, where it is particularly difficult to identify YSOs at both infrared and submillimeter wavelength regimes. For this reason, we used the sum within a 20\arcsec\ aperture on \Herschel/PACS 70~\micron\ maps centered on the locations of these objects to measure an upper limit on their respective 70~\micron\ fluxes. These upper limits (shown in Figures~\ref{fig:seds} and \ref{fig:seds2}) are consistent with the 70~\micron\ fluxes of YSOs detected with \Herschel/PACS, so we could not confirm or rule out whether these sources are indeed protostellar. Furthermore, the average 70~\micron\ flux in this area is $\sim$0.7~Jy, which would adequately conceal compact emission from YSOs, as the average flux of \Herschel/PACS--identified YSOs at 70~\micron\ is $\sim$1~Jy. We therefore continue to refer to these sources as candidate protostellar objects based on the detection of compact emission at submillimeter wavelengths. 

SCUBA-2 fluxes were used to measure the masses of the circumstellar material (disk + envelope) where the envelope has a larger contribution to the total mass for earlier-class objects. We assumed a temperature of 20 K and a $\beta$ value of 1 for the mass calculation to facilitate comparison with YSO populations in other clouds. We verified the assumption of these parameters to represent the coolest dust by performing a $\chi$-squared minimization of the SED at long wavelengths for the 19 robust YSOs in our sample that have $\geq 3\sigma$ detections from 160 to 850~\micron. The average fitted temperature (17 $\pm$ 6 K) and the average fitted $\beta$ value (1.1 $\pm$ 0.6) are consistent with the fiducial parameters assumed for such mass measurements and exhibit a reasonable representation of our YSO population. Furthermore, the predicted curve of the SED at submillimeter wavelengths that results from assuming these fiducial values agrees well with the observed shape of the SED. 

The resulting circumstellar mass distribution reflects that we are sensitive to the high end of the mass distribution relative to other measured populations. The circumstellar masses measured here are generally higher than the circumstellar masses measured in other Gould Belt clouds. For the more nearby GB clouds, specifically Taurus and $\rho$ Oph \citep{AndrewsWilliams2005,AndrewsWilliams2007}, these regions are also observed with SCUBA-2. The distance discrepancy results in more material (both from the outer circumstellar envelope and from the extended cloud emission) being within the beam for Auriga--Cal. The YSOs measured in Orion A, and therefore at a similar distance as the YSOs in Auriga--Cal, have been measured with an interferometer \citep{MannWilliams2010,Mannetal2014,Mannetal2015}; thus, these fluxes also have less contribution from the larger-scale emission (which has been spatially filtered out). The $\sigma$ Orionis cluster was also observed with SCUBA-2 \citep{Williamsetal2013}; however, this cluster is much older, so the YSOs within in it are generally more evolved and therefore have less circumstellar matter.

Finally, we compared the ratios of YSOs in Auriga--Cal and Orion A. There are 11 times more \Spitzer-identified YSOs in Orion A than in Auriga--Cal within the SCUBA-2 coverage of the two clouds in the JCMT GBS and 7 times more submillimeter compact sources. The similarity between these ratios shows that the disparity of star formation populations between the clouds observed in the infrared extends to the submillimeter. These ratios also suggest that the relative star formation rates of the two clouds have not varied over the Class II lifetime and suggests that Auriga--Cal will maintain a paltry population of stellar objects with respect to Orion A.


\acknowledgments

HBF and BCM acknowledge a Discovery Grant from the Natural Science \& Engineering Research Council (NSERC) of Canada.
HBF acknowledges support from the Alfred Bader Scholarship in Memory of Jean Royce administered by Queen's University, Canada.
The James Clerk Maxwell Telescope has historically been operated by the Joint Astronomy Centre on behalf of the Science and Technology Facilities Council of the United Kingdom, the National Research Council of Canada, and the Netherlands Organisation for Scientific Research.  Additional funds for the construction of SCUBA-2 were provided by the Canada Foundation for Innovation.
This research used the services of the Canadian Advanced Network for Astronomy Research (CANFAR), which in turn is supported by CANARIE, Compute Canada, the University of Victoria, the National Research Council of Canada, and the Canadian Space Agency. This research used the facilities of the Canadian Astronomy Data Centre, operated by the National Research Council of Canada with the support of the Canadian Space Agency. 
This research also made use of APLpy \citep{aplpy2012}, an open-source plotting package for Python hosted at http://aplpy.github.com

\bibliographystyle{apj}
\bibliography{SCUBA2-Auriga-Cal-astroph.bbl} 

\section{Appendix: Identified Sources}

Here we include an appendix of the full set of images for the compact sources identified with \getsources, not including the four examples shown in Figure~\ref{fig:vet}. These are shown in Figure~\ref{fig:appendix}.

\clearpage
\begin{figure*}[h]
\includegraphics[trim=2.25cm 3.5cm 0cm 5.5cm, clip=True, width=7.5in,page=2]{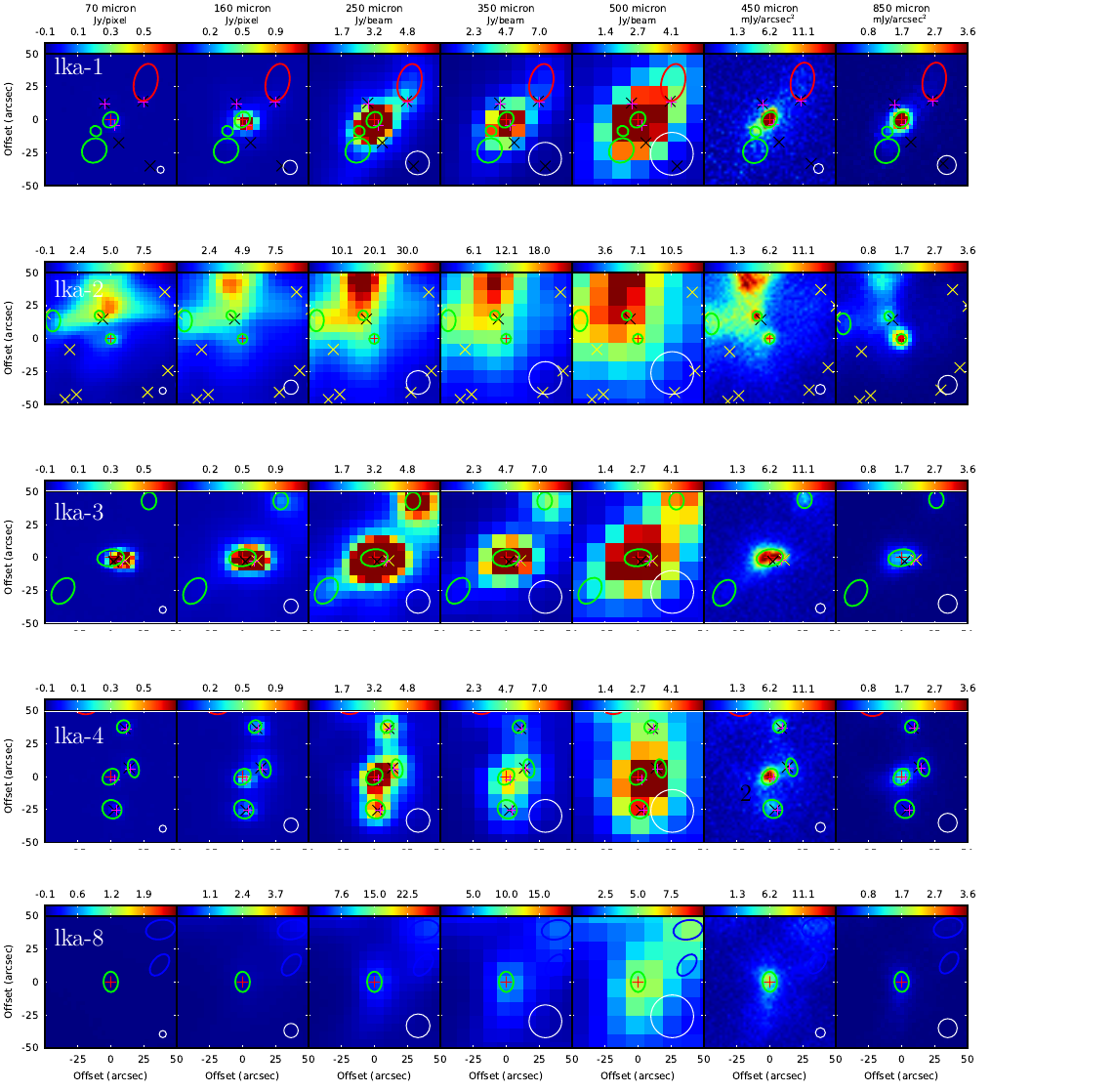}
\caption{\footnotesize{Quality-assurance maps for compact sources extracted using \getsources\ showing (left to right) the \Herschel/PACS (70 and 160~\micron), \Herschel/SPIRE (250, 350, and 500~\micron), and SCUBA-2 (450 and 850~\micron) maps. Each panel is centered on the compact source in question, which is marked with crosshairs. Elliptical regions are the same as in Figures~\ref{fig:pongs} with blue, red, and green ellipses marking identified candidate YSOs at 450~\micron, 850~\micron, or both, respectively, and with major and minor FWHM and orientation according to the source properties measured with \getsources. The internal \getsources\ ID is listed in the upper left corner of the 70~\micron\ map for each source. (We use the internal \getsources\ ID to identify the compact sources, as we make the plots for the complete set of 79 compact sources and not just the subset of 59 candidate YSOs.) This figure is the extended version of Figure 4.
}\label{fig:appendix}}
\end{figure*}

\begin{figure*}[h]
\ContinuedFloat
\includegraphics[trim=2.25cm 3.5cm 0cm 5.5cm, clip=True, width=7.5in,page=3]{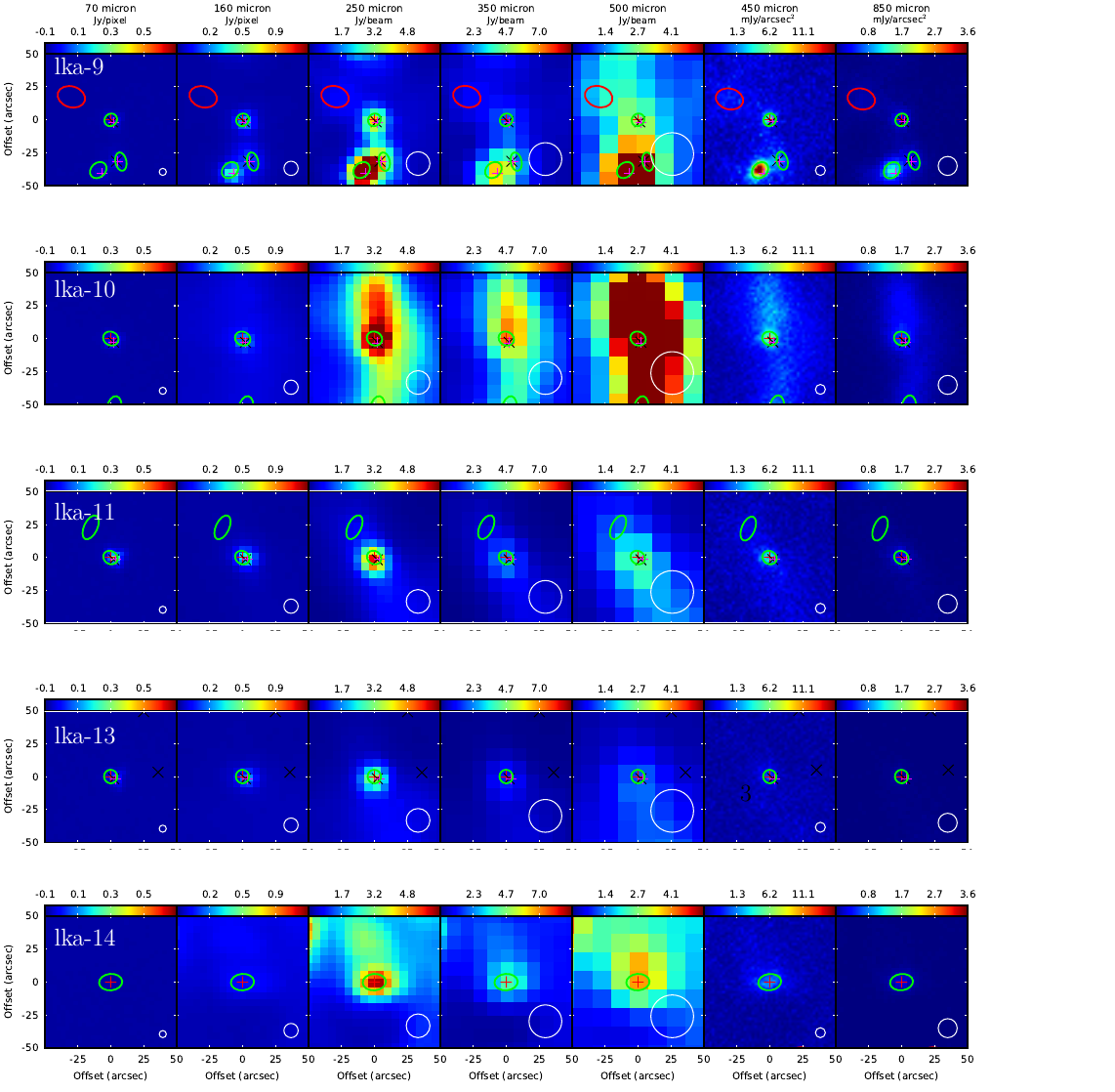}
\caption{\small{ 
}\label{fig:appendix}}
\end{figure*}

\begin{figure*}[h]
\ContinuedFloat
\includegraphics[trim=2.25cm 3.5cm 0cm 5.5cm, clip=True, width=7.5in,page=4]{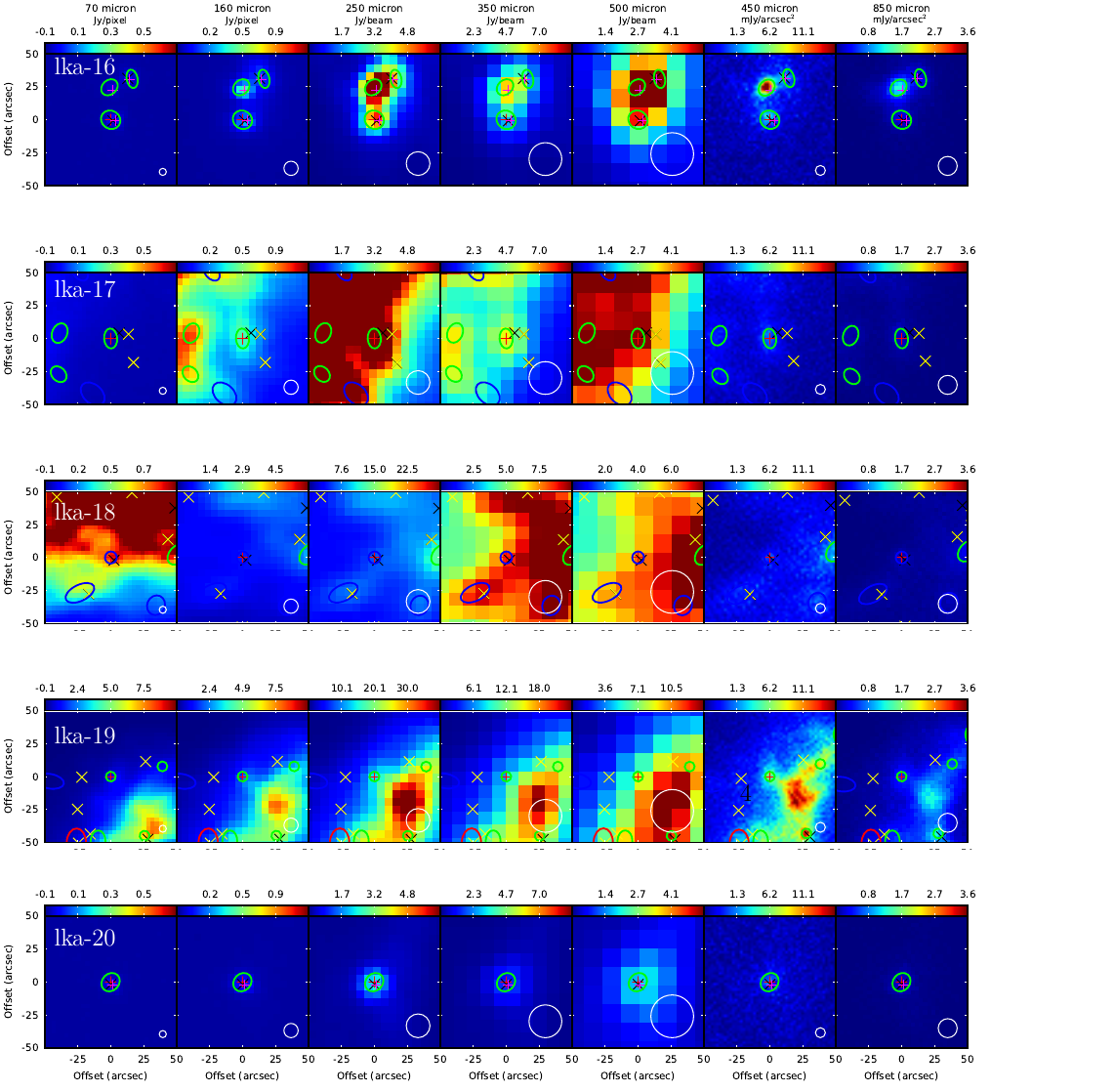}
\caption{\small{ 
}\label{fig:appendix}}
\end{figure*}

\begin{figure*}[h]
\ContinuedFloat
\includegraphics[trim=2.25cm 3.5cm 0cm 5.5cm, clip=True, width=7.5in,page=5]{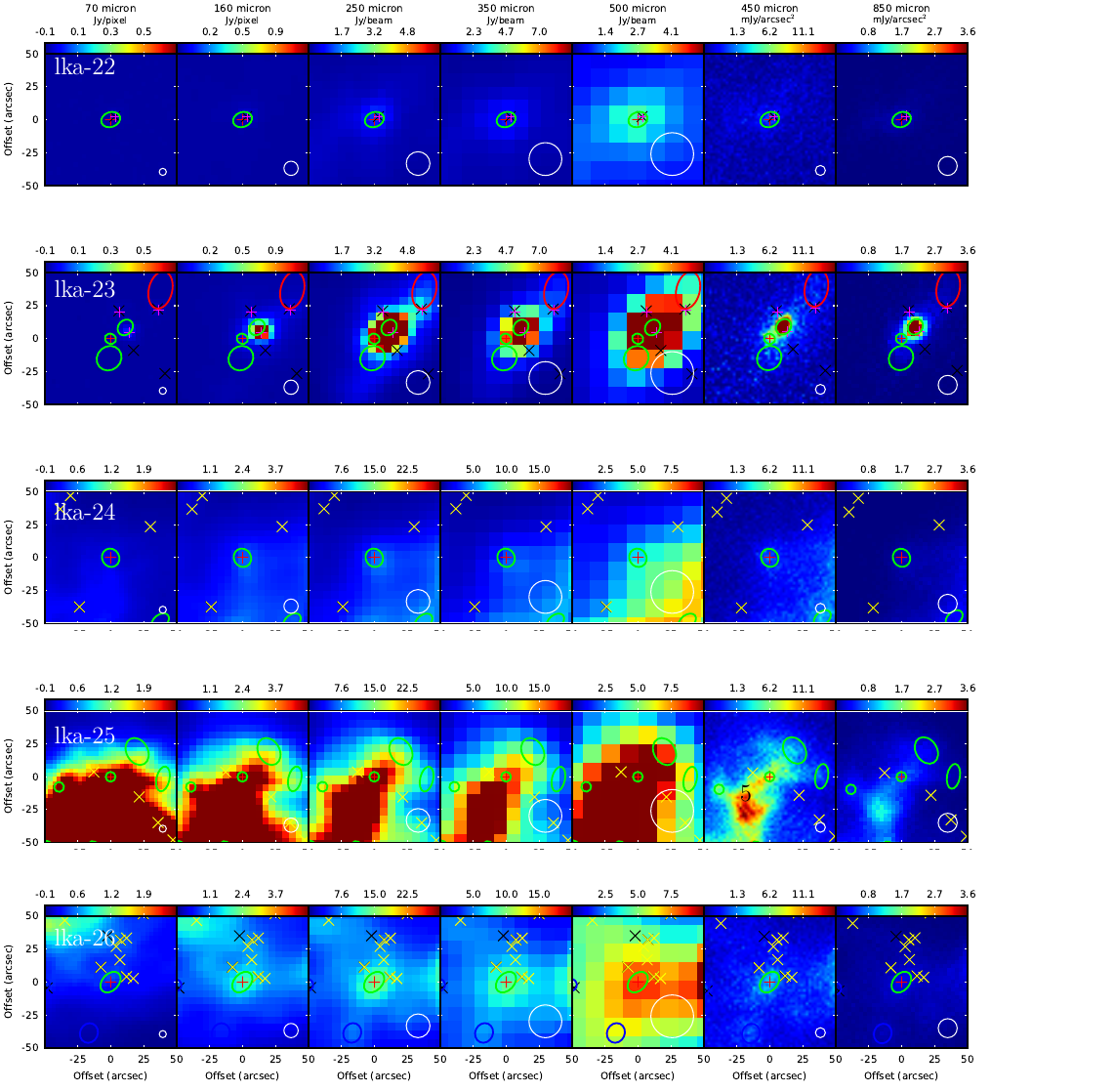}
\caption{\small{ 
}\label{fig:appendix}}
\end{figure*}

\begin{figure*}[h]
\ContinuedFloat
\includegraphics[trim=2.25cm 3.5cm 0cm 5.5cm, clip=True, width=7.5in,page=6]{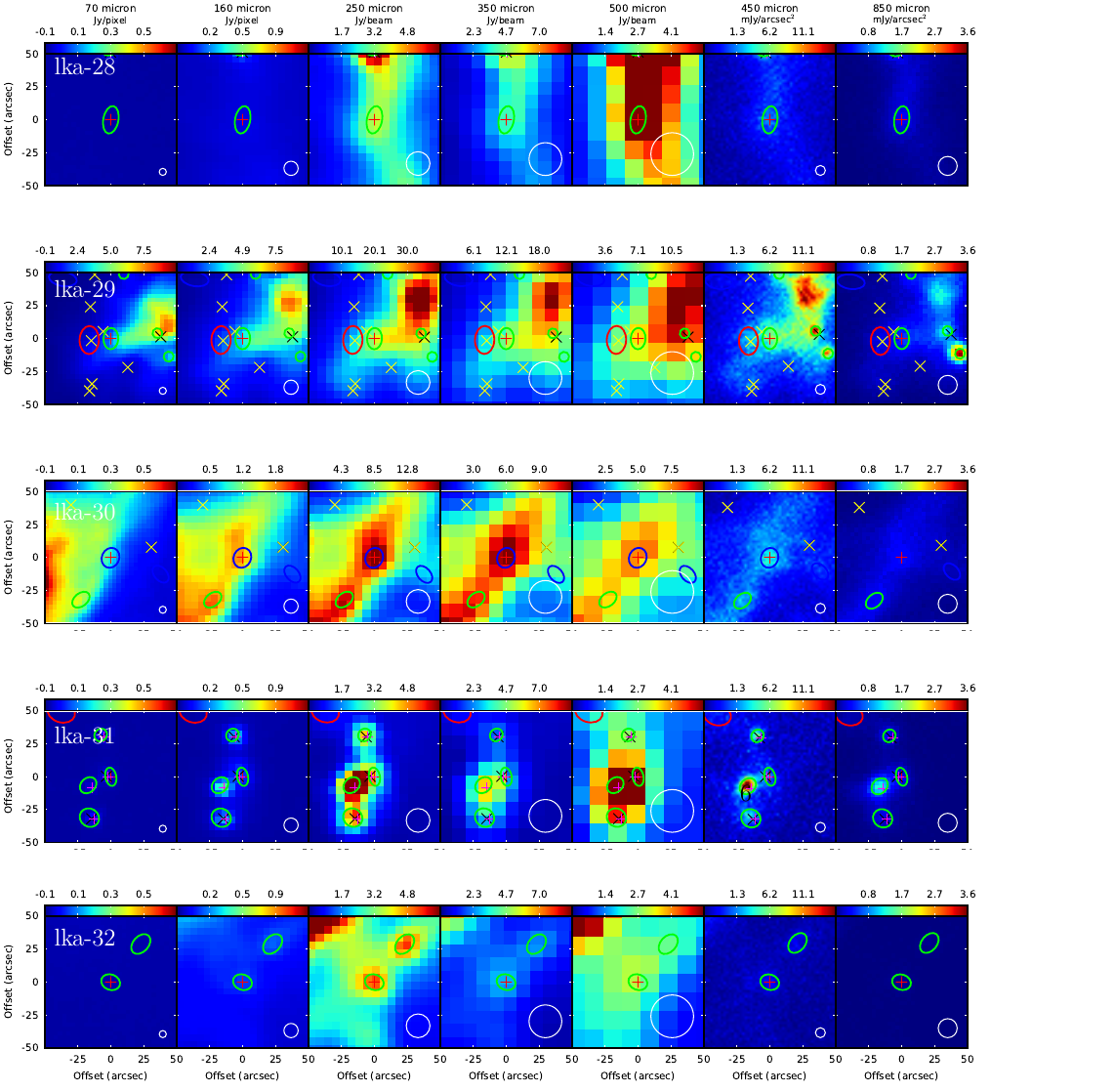}
\caption{\small{ 
}\label{fig:appendix}}
\end{figure*}

\begin{figure*}[h]
\ContinuedFloat
\includegraphics[trim=2.25cm 3.5cm 0cm 5.5cm, clip=True, width=7.5in,page=7]{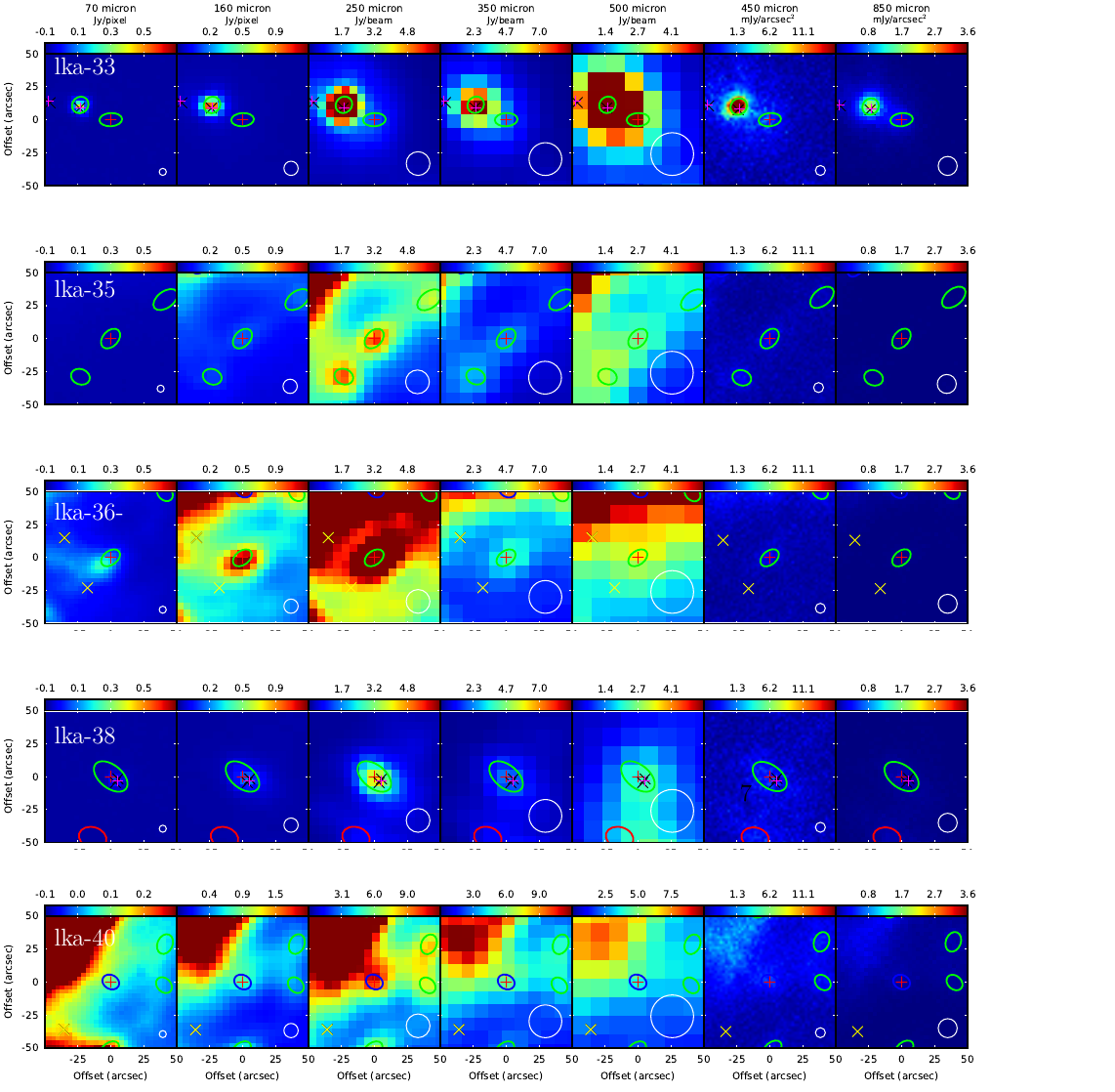}
\caption{\small{ 
}\label{fig:appendix}}
\end{figure*}

\begin{figure*}[h]
\ContinuedFloat
\includegraphics[trim=2.25cm 3.5cm 0cm 5.5cm, clip=True, width=7.5in,page=8]{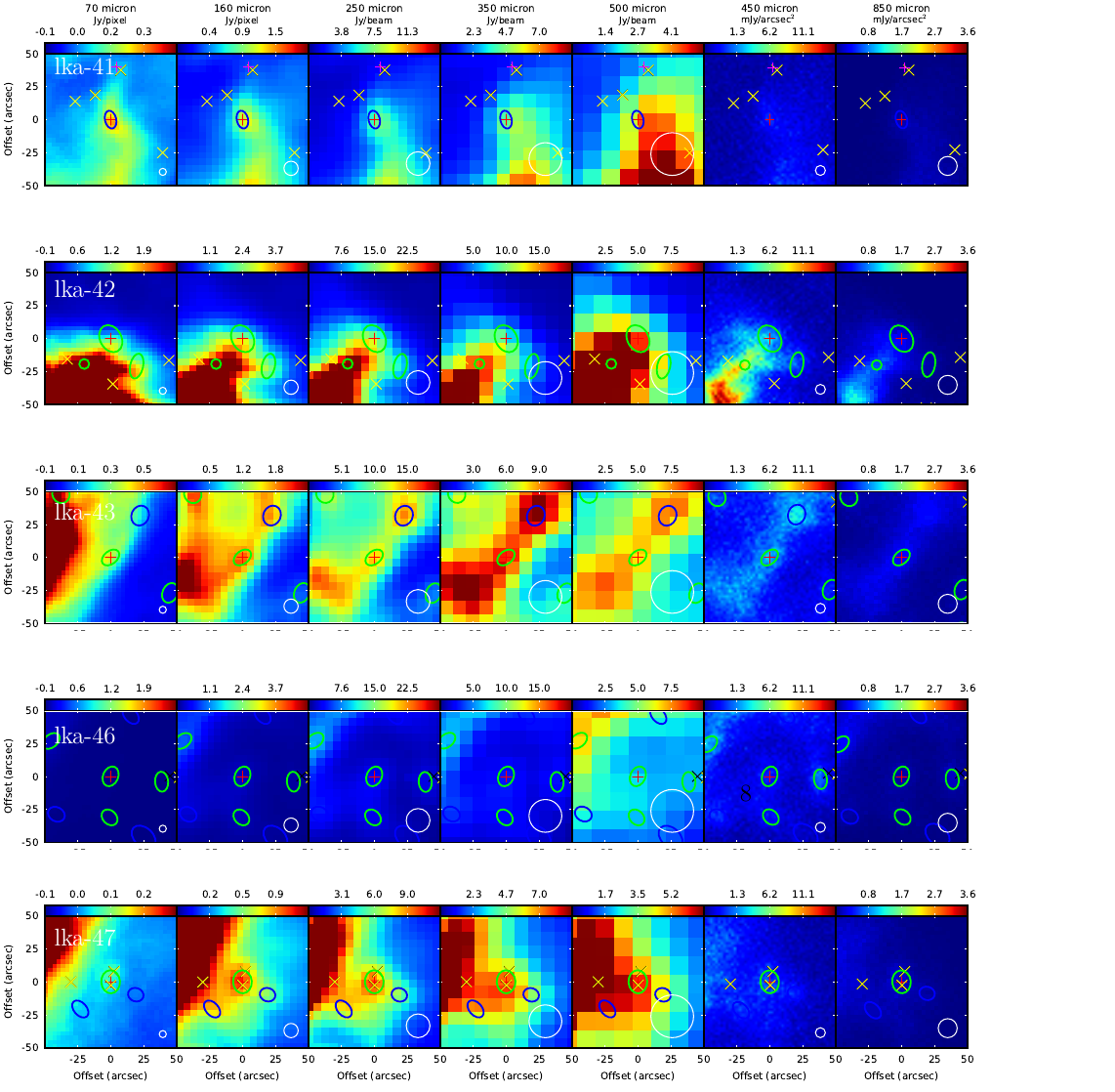}
\caption{\small{ 
}\label{fig:appendix}}
\end{figure*}

\begin{figure*}[h]
\ContinuedFloat
\includegraphics[trim=2.25cm 3.5cm 0cm 5.5cm, clip=True, width=7.5in,page=9]{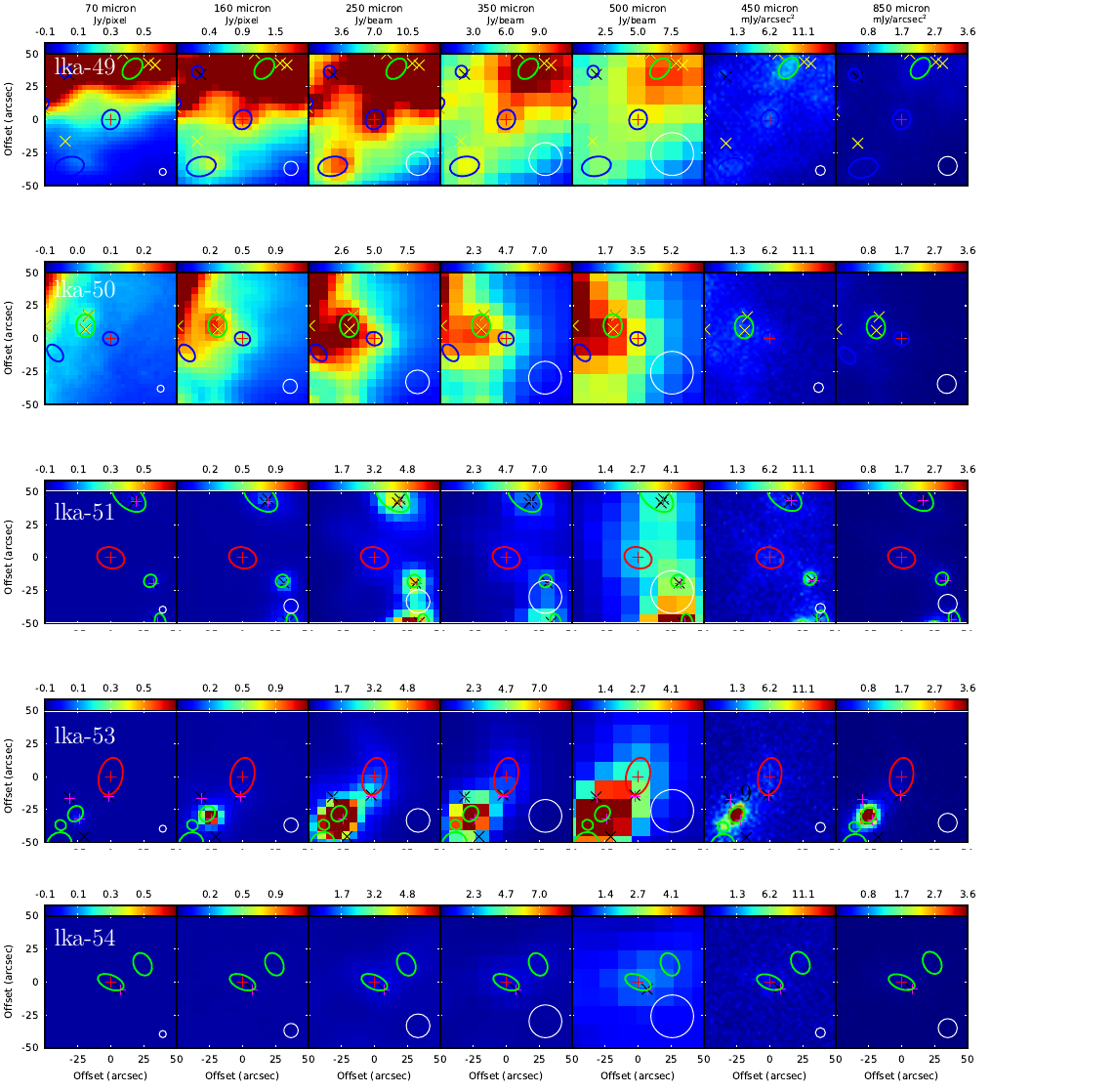}
\caption{\small{ 
}\label{fig:appendix}}
\end{figure*}

\begin{figure*}[h]
\ContinuedFloat
\includegraphics[trim=2.25cm 3.5cm 0cm 5.5cm, clip=True, width=7.5in,page=10]{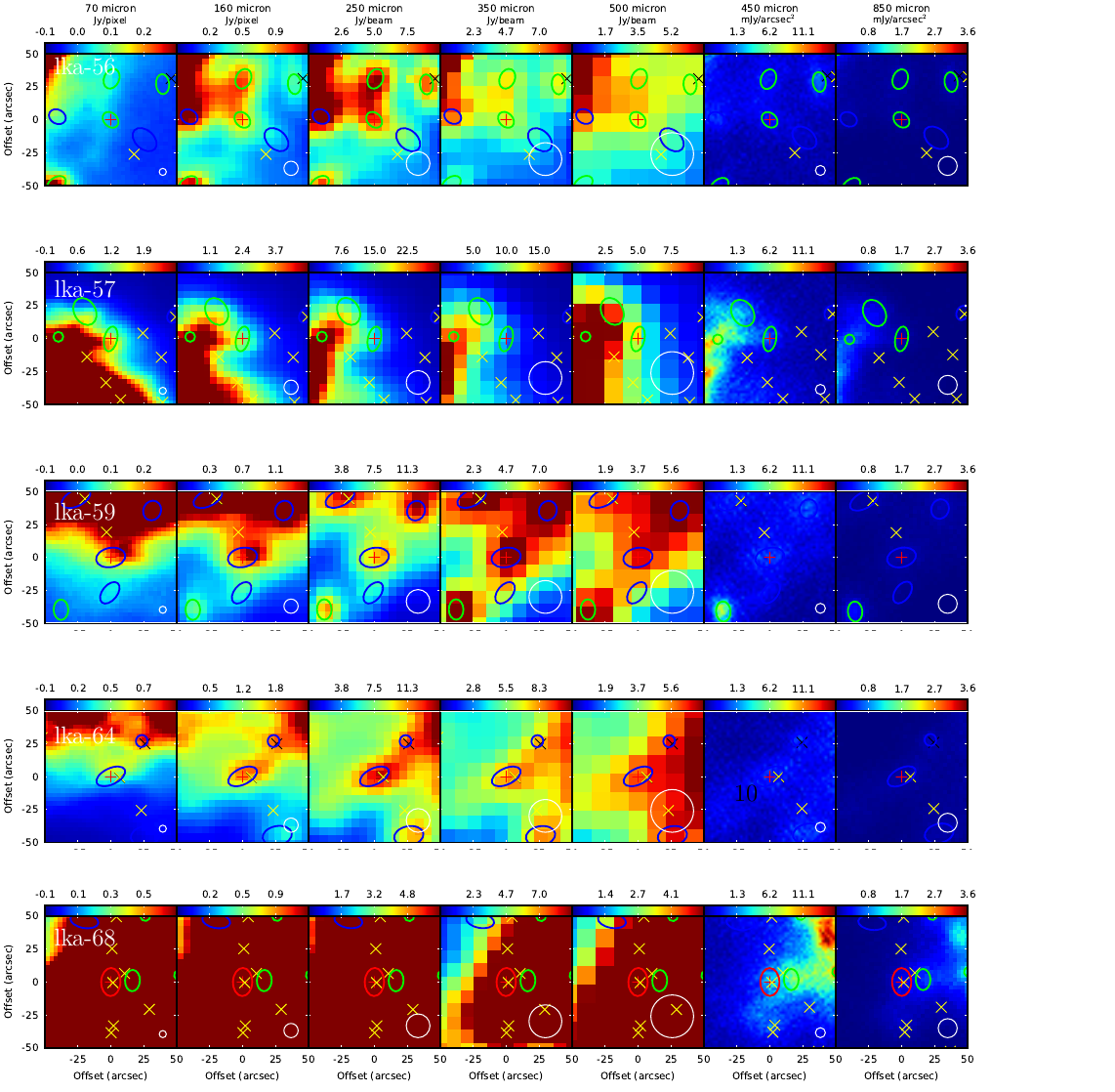}
\caption{\small{ 
}\label{fig:appendix}}
\end{figure*}

\begin{figure*}[h]
\ContinuedFloat
\includegraphics[trim=2.25cm 3.5cm 0cm 5.5cm, clip=True, width=7.5in,page=11]{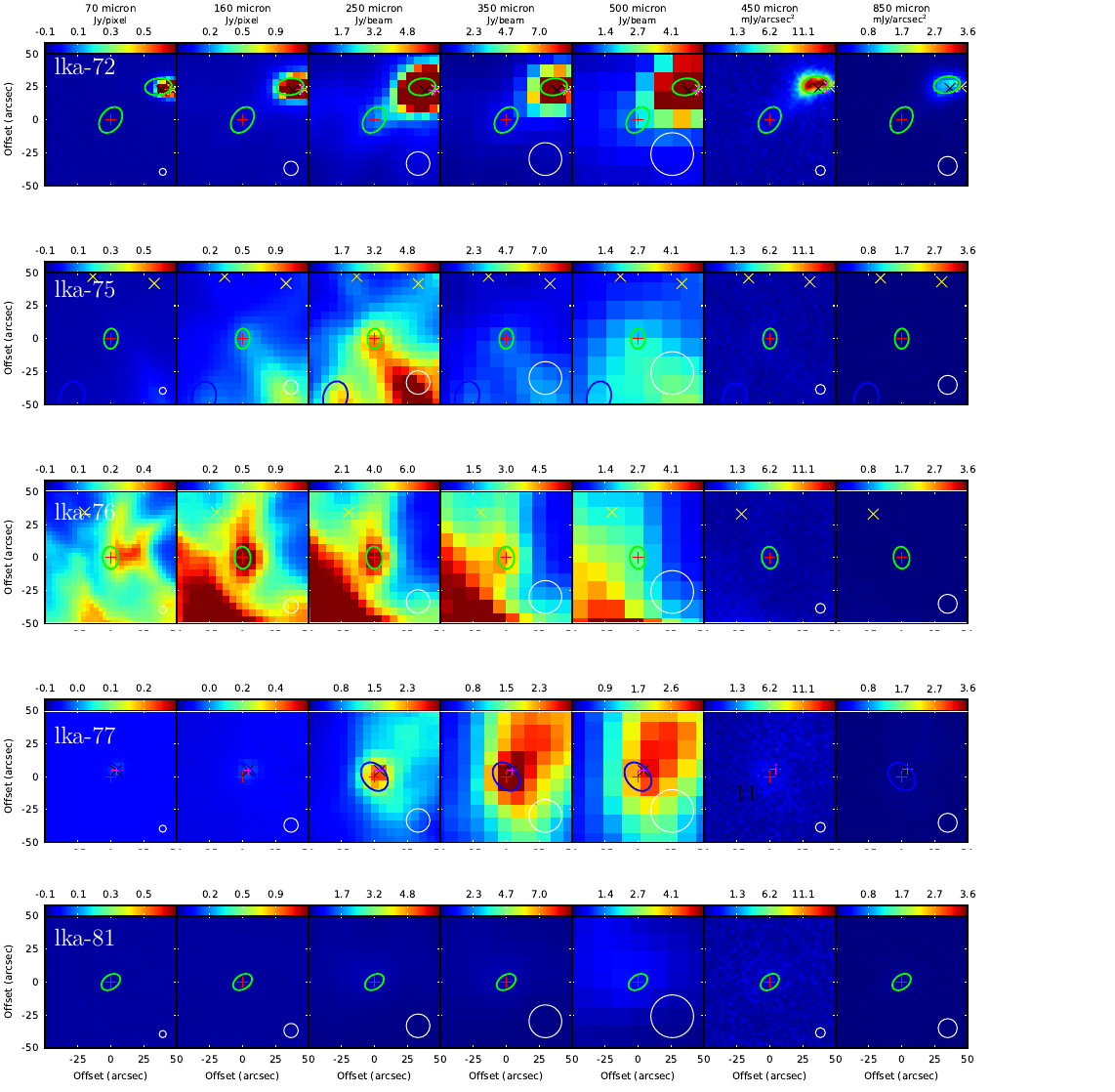}
\caption{\small{ 
}\label{fig:appendix}}
\end{figure*}

\begin{figure*}[h]
\ContinuedFloat
\includegraphics[trim=2.25cm 3.5cm 0cm 5.5cm, clip=True, width=7.5in,page=12]{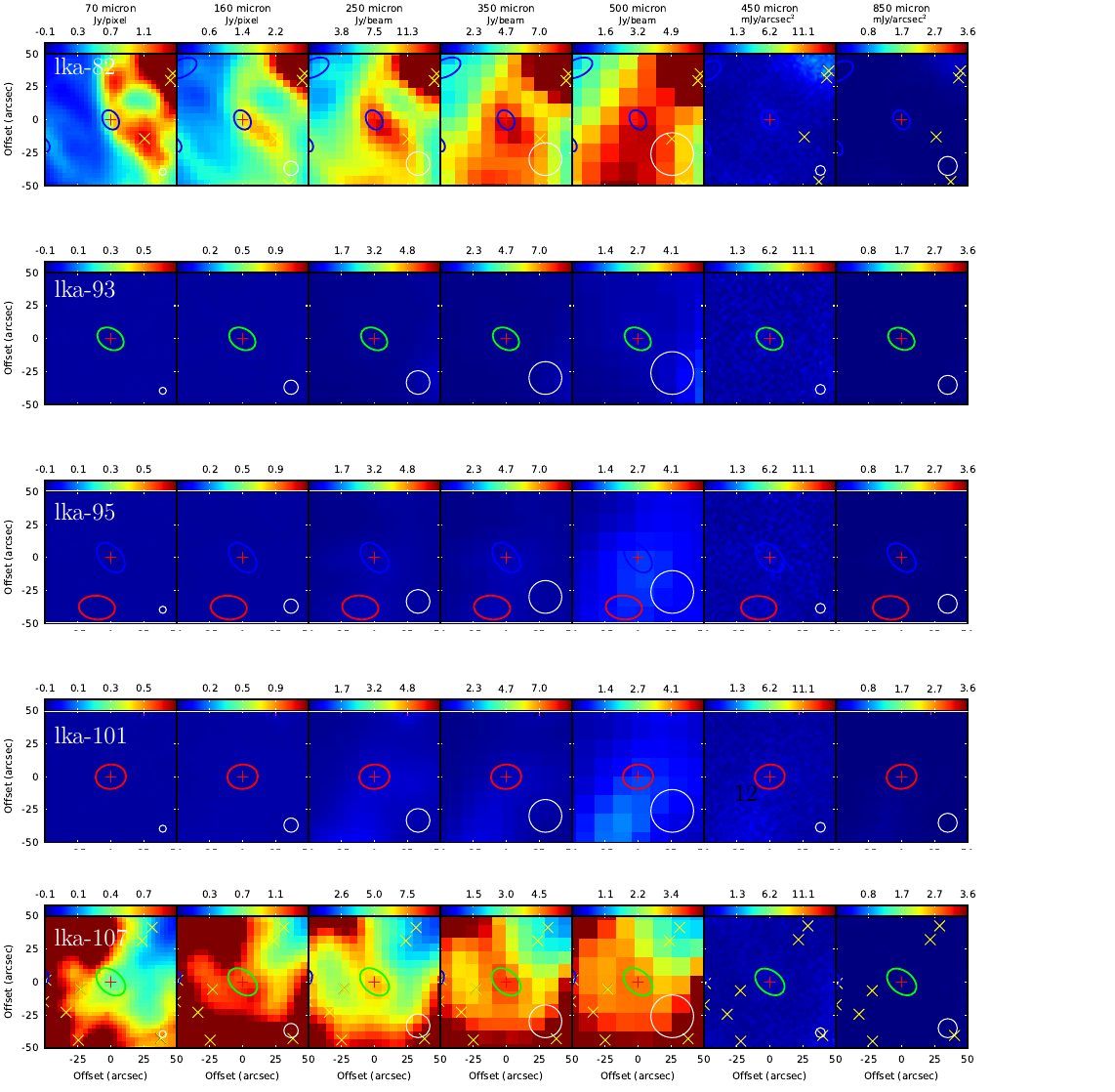}
\caption{\small{ 
}\label{fig:appendix}}
\end{figure*}

\begin{figure*}[h]
\ContinuedFloat
\includegraphics[trim=2.25cm 3.5cm 0cm 5.5cm, clip=True, width=7.5in,page=13]{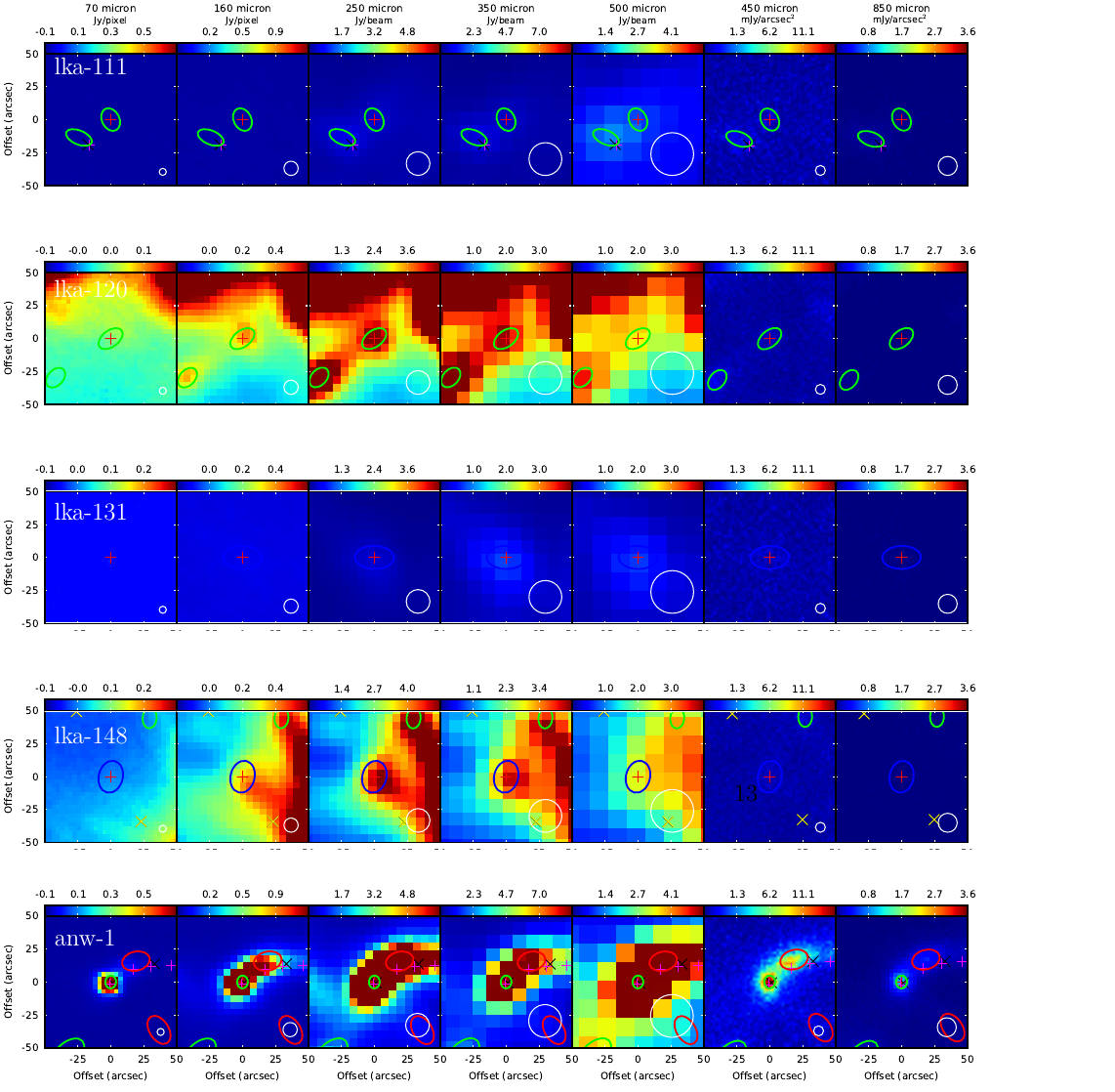}
\caption{\small{ 
}\label{fig:appendix}}
\end{figure*}

\begin{figure*}[h]
\ContinuedFloat
\includegraphics[trim=2.25cm 3.5cm 0cm 5.5cm, clip=True, width=7.5in,page=14]{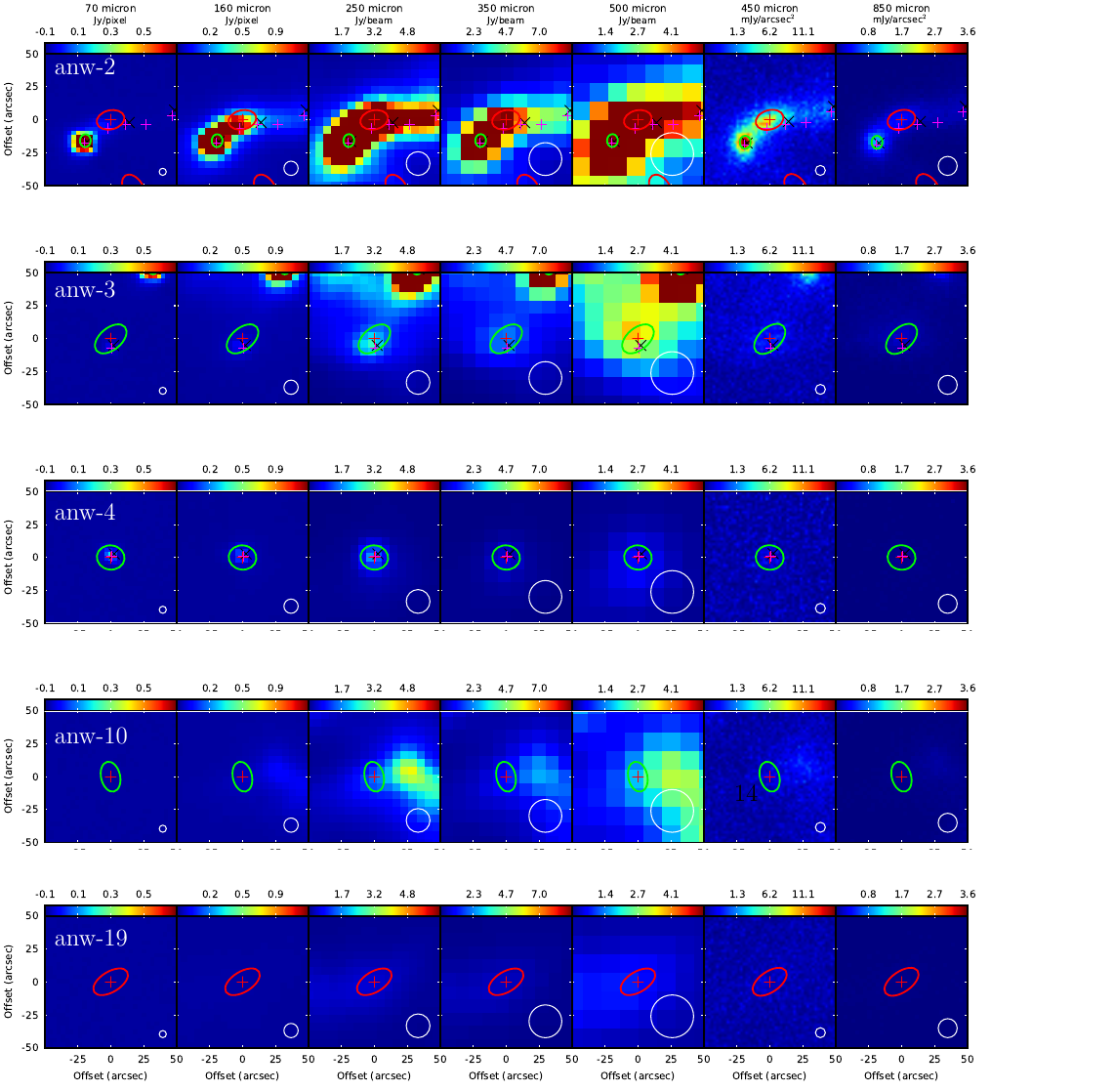}
\caption{\small{ 
}\label{fig:appendix}}
\end{figure*}

\begin{figure*}[h]
\ContinuedFloat
\includegraphics[trim=2.25cm 3.5cm 0cm 5.5cm, clip=True, width=7.5in,page=15]{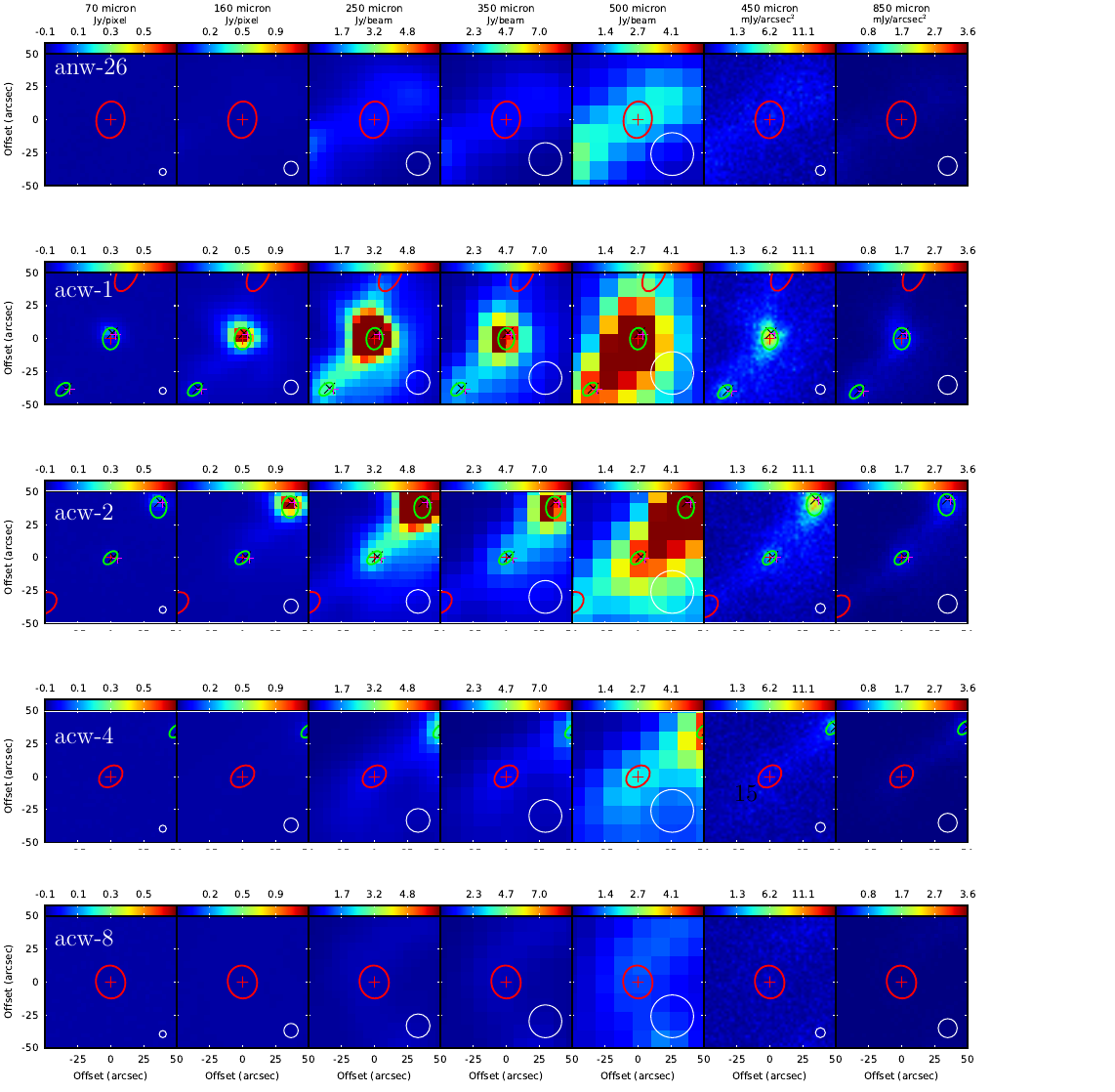}
\caption{\small{ 
}\label{fig:appendix}}
\end{figure*}

\begin{figure*}[h]
\ContinuedFloat
\includegraphics[trim=2.25cm 3.5cm 0cm 5.5cm, clip=True, width=7.5in,page=16]{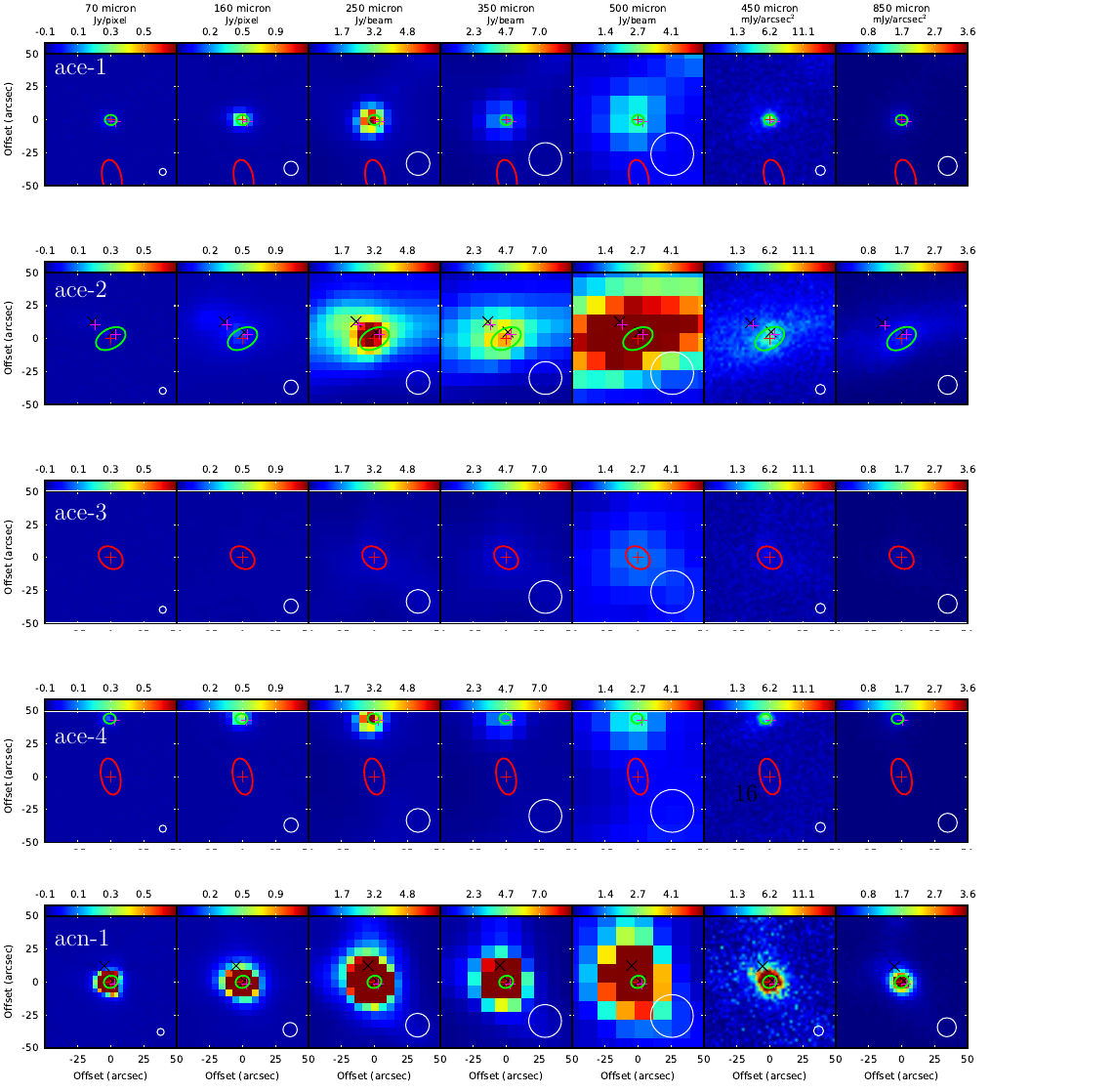}
\caption{\small{ 
}\label{fig:appendix}}
\end{figure*}

\end{document}